\documentclass[preprint]{elsarticle}

\usepackage{graphicx}
\usepackage{mathrsfs}
\usepackage{dcolumn}
\usepackage{dashrule}
\usepackage{amsmath}
\usepackage{multirow}
\usepackage{stmaryrd}
\usepackage{bm}
\usepackage[hmargin=2cm,vmargin=2cm]{geometry}

\newcommand\Rey{\mbox{\textit{Re}}}
\newcommand\Ca{\mbox{\textit{Ca}}}

\newcommand\bnabla{\boldsymbol{\nabla}}
\newcommand\bcdot{\boldsymbol{\cdot}}

\newcommand{\RN}[1]{%
  \textup{\uppercase\expandafter{\romannumeral#1}}%
}
\DeclareMathOperator{\Res}{Res}

\usepackage{multirow}
\usepackage{booktabs}

\begin{document}

\begin{frontmatter}

\title{Theoretical Model of a Finite Force at the Moving Contact Line}

\author[mymainaddress]{Peter Zhang}

\author[mymainaddress,mysecondaryaddress]{Kamran Mohseni\corref{mycorrespondingauthor}}
\cortext[mycorrespondingauthor]{Corresponding author}
\ead{mohseni@ufl.edu}

\address[mymainaddress]{Department of mechanical and aerospace, University of Florida, Gainesville, FL.}
\address[mysecondaryaddress]{Department of Electrical and Computer Engineering, University of Florida, Gainesville, FL.}

\begin{abstract}
In theoretical analyses of the moving contact line, an infinite force along the solid wall has been reported based off the non-integrable stress along a {\it single} interface. In this investigation we demonstrate that the stress singularity is integrable and results in a {\it finite} force at the moving contact line if the contact line is treated as a one-dimensional manifold and all {\it three} interfaces that make up the moving contact line are taken into consideration. This is due to the dipole nature of the vorticity and pressure distribution around the moving contact line. Mathematically, this finite force is determined by summing all the forces that act over an infinitesimally small cylindrical control volume that encloses the entire moving contact line. With this finite force, we propose a new dynamic Young's equation for microscopic dynamic contact angle that is a function of known parameters only, specifically the interface velocity, surface tension, and fluid viscosity. We combine our model with Cox's model for apparent dynamic contact angle and find good agreement with published dynamic contact angle measurements.
\end{abstract}

\begin{keyword}
moving contact line, dynamic contact angle, multiphase flows
\end{keyword}

\end{frontmatter}

\section{Introduction}

The moving contact line (MCL) is a unique and challenging problem that influences many natural and industrial processes such as drop impact \cite{YarinAL:06a}, boiling \cite{DhirVK:98a}, industrial coatings \cite{WeinsteinS:04a}, and inkjet printing \cite{DerbyB:10a}, among others. Some of the aspects that make the MCL such a complex problem include its multiscale nature \cite{SnoeijerJH:13b}, the apparent breakdown of the no-slip assumption \cite{Dussan:76a}, hysteresis \cite{EralH:13a}, and dependence on surface properties \cite{QuereD:08a}. In recent years industrial and medical applications have spurred MCL research towards even more challenging problems such as wetting failure \cite{deGennesPG:85a,LandauL:88a}, air entrainment \cite{KumarS:14a}, electrowetting \cite{MugeleF:05a}, and phase change at the contact line \cite{SnoeijerJH:12a,SnoeijerJH:13a}. While these problems are important for advancing multiphase technology, there is still no consensus on the physics that govern contact line movement over a smooth surface \cite{SnoeijerJH:13b,SpeltP:14a,BlakeTD:06a,BonnD:09a}. Understanding the MCL dynamics over a smooth surface is essential to developing models for more complex wetting phenomena and thus it has remained a topic of significant interest for several decades.

\begin{figure}
\centering
\begin{minipage}{0.99\linewidth}\begin{center}
 \includegraphics[width=0.49\linewidth]{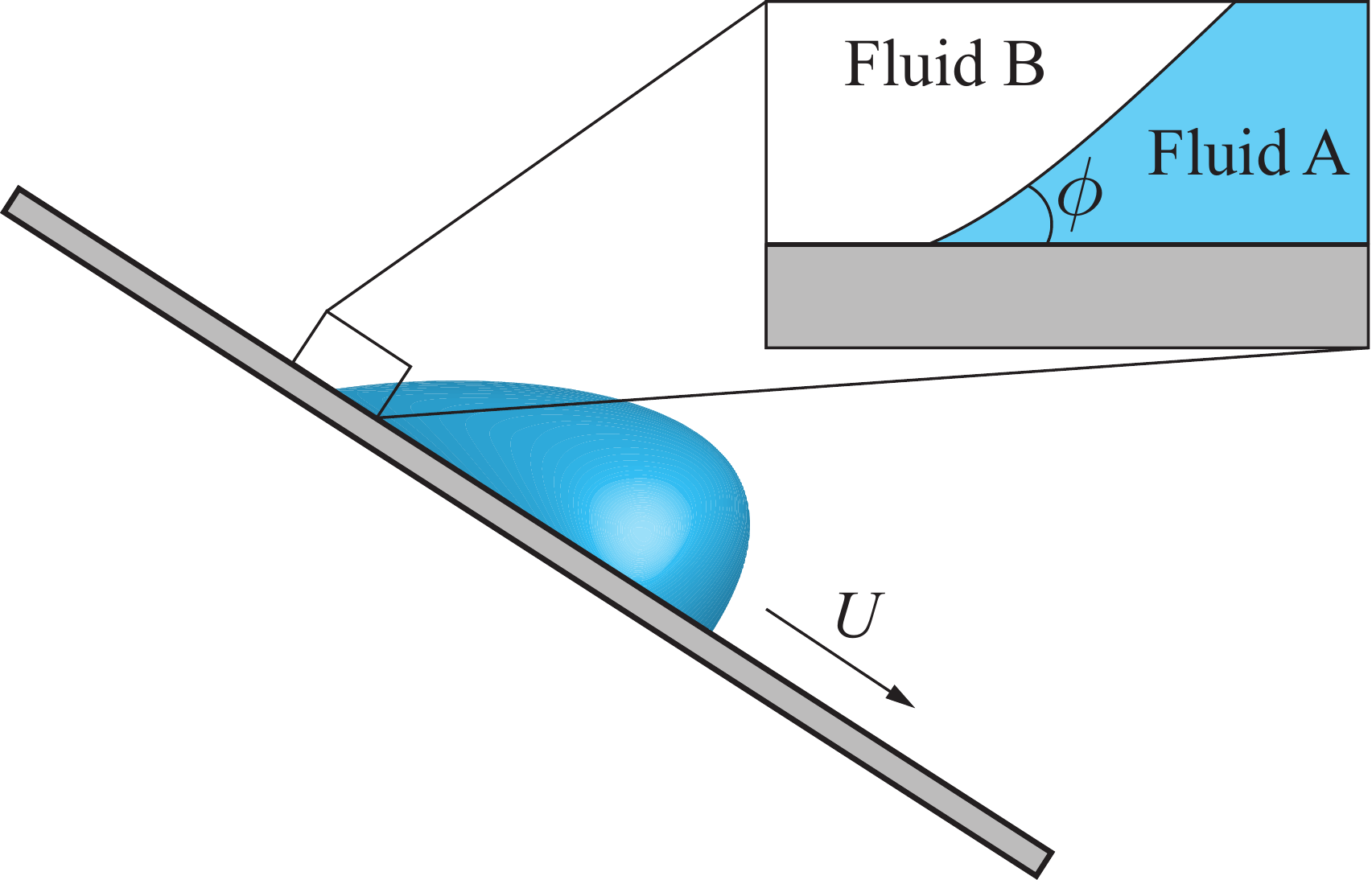}
\end{center}\end{minipage}
\caption{Schematic of a droplet sliding down an inclined plate. Near the moving contact line, the interface has minimal curvature and intersects the solid with a dynamic contact angle $\phi$.}
\label{fig:schematic}
\end{figure}

Two early investigations of this fundamental problem were conducted by Moffatt \cite{MoffattHK:64a} and Huh \& Scriven \cite{HuhC:71a}. In these works, the MCL geometry was modeled by three planar interfaces intersecting with a contact angle $\phi$ and characterized by an interface velocity $U$, as shown in figure~\ref{fig:schematic}. Moffatt examined a viscous fluid displacing an inviscid gas and reported a solution with an ``infinite stress and pressure (both of order $r^{-1}$) on the the plate at the corner''. Huh \& Scriven extended this analysis to two viscous fluids and reported that ``the total force exerted on the solid surface is logarithmically infinite" due to the singular stress at the MCL. These investigations, among others \cite{HockingLM:77a,deGennesPG:85a,CoxRG:86a,ShikhmurzaevYD:97a,EggersJ:04a,BlakeTD:06a,BonnD:09a,SnoeijerJH:13b,SpeltP:14a}, have concluded that the hydrodynamic solution does not accurately model the MCL because an infinite force is not physical and this result has come to be known as the MCL problem, or ``Huh \& Scriven's paradox'' \cite{BonnD:09a}.

Since these early works, several MCL theories have been developed and include the molecular kinetic theory (MKT) \cite{BlakeTD:69a}, interface formation theory (IFT) \cite{ShikhmurzaevYD:97a,ShikhmurzaevYD:07a}, and hydrodynamic theory \cite{CoxRG:86a,VoinovOV:76a,Dussan:76a} among others \cite{PetrovP:92a,PismenL:02a,SeppecherP:96a}. In many of these models, microscopic physical mechanisms relieve the stress singularity and allows the interface to move relative to the solid without inducing a singular force. MKT describes the motion of the contact line as a molecular process where molecules attach and detach from the solid surface with some characteristic frequency and displacement. The rate at which these molecules are displaced induces changes in the local surface tension which subsequently affects the dynamic contact angle. Typically the length scale of the molecular displacement is on the order of nanometers while the frequency of this displacement is inversely proportional to the viscosity of the fluid. Through experiment the molecular frequency and displacement have been determined for specific fluid-solid pairs however, it is often difficult to predict these parameters for general cases \cite{BlakeTD:06a}.

Interface formation theory is built on the premise that near a MCL, fluid elements move from one interface to another in finite time so that the fluid exhibits a rolling type motion rather than a sliding type motion associated with fluid slip. As fluid elements transition from one interface to another, IFT posits that the fluid properties will gradually change such that a surface tension gradient will be generated near the MCL. Shikhmurzaev \cite{ShikhmurzaevYD:97a} proposes an equation of state to relate the surface tension to the interfacial density multiplied by a phenomenological coefficient. Assuming that the Young's equation is valid for dynamic contact lines, the predicted surface tension gradient results in a change in contact angle. Proponents of IFT report that the model contains no singularities, preserves the rolling kinematics of MCLs, and captures the effects of the local flow field on the dynamic contact angle \cite{ShikhmurzaevYD:06a}. Based on these characteristics, and the possibility of simulating high Capillary number flows without slip, IFT has been praised as a potentially far-reaching approach \cite{BlakeTD:06a}.

The hydrodynamic theory of the MCL is based on classical continuum fluid mechanics but relaxes the no-slip condition. In many cases the Navier-slip boundary condition \cite{NavierCL:23a} and a constant slip length is used to determine the flow near the MCL. However, recent molecular dynamics simulations have indicated that the slip length is dependent on the fluid stress \cite{Troian:97a} and that fluid slip near the MCL may require a more generalized slip boundary condition \cite{Mohseni:16b}. In general, hydrodynamic models find that the interface shape varies logarithmically with respect to distance from the contact line and is a function of the microscopic contact angle and inner to outer length scale ratio \cite{CoxRG:86a,VoinovOV:76a,HockingLM:82a}. In many applications, a constant microscopic contact angle and length scale ratio allows the hydrodynamic model to capture the shape of the interface \cite{RameE:96a,SuiY:13a}. However, there is no physical reason why these parameters should be constant and there have been a growing number of publications that have reported that these parameters should vary as a function of contact line velocity \cite{RameE:04a,ShenC:98a,ZhouM:92a}. 

In each of the theories reviewed above, molecular scale physics are used to remove the stress singularity at the MCL. As a consequence, the MCL becomes a multiscale problem that couples macroscopic dynamics to microscopic physical parameters like the molecular equilibrium frequency of MKT, the interfacial density of IFT, or the slip length in hydrodynamic models. Currently these microscopic parameters are difficult to determine theoretically and are most often obtained by fitting the theory to experimental measurements. Through the fitting process, most of these theories have reported agreement with experimental measurements \cite{HoffmanRL:75a,CoxRG:86a,ShikhmurzaevYD:02a,SevenoD:09a,BlakeT:11a,ItoT:15a} and it has been difficult to judge which theory captures the true physics of the contact line. As a result, the physics of the MCL are still debated to this day and the field continues to grow larger and more diverse. For additional information regarding the MCL, we refer the reader to the following references \cite{BlakeTD:06a,Dussan:79a,BonnD:09a,SnoeijerJH:13b,SpeltP:14a,deGennesPG:85a}. 

Looking back on the evolution of the MCL problem, it appears that many investigations were motivated by the conclusions of Moffatt and Huh \& Scriven who determined that a stress that scales as $1/r$ is not integrable and that the hydrodynamic solution of the MCL subject to the no-slip boundary condition incorrectly predicts an infinite force. Interestingly we have observed that a continuum field that scales as $1/r$ is treated as an {\it integrable} singularity in other fields, and even in continuum fluid flows. In electromagnetism, an electric field that scales as $1/r$ is integrable and correlated to the total charge \cite{GriffithsDJ:72a}. In potential flow theory, a velocity field that scales with $1/r$ is integrable and directly related to the total mass flux of a line source \cite{Batchelor:67a}. Even in Stokes flows, the two-dimensional Stokeslet contains an stress singularity that scales as $1/r$ and is considered an integrable singularity that is correlated to a finite force \cite{CrowdyDG:10a}. Given these numerous examples of integrable $1/r$ singularities, one begins to wonder why is the MCL different?

Motivated by these observations, this work will revisit the classic hydrodynamic solution and present an alternative perspective of the stress singularity. We will begin with a brief review of the Stokes solution to the MCL problem in \S\ref{sect:geo_sol}.  In \S\ref{sect:f_MCL}, we will show that classical hydrodynamics models the contact line region as a mathematical line, i.e. a one-dimensional manifold. By treating the MCL as a one-dimensional manifold, we find that the total force exerted by the fluid is finite and a function of the surface tension, interface velocity, and fluid viscosity. \S\ref{sect:discussion} presents a discussion of this finite force and a comparison with previous works to show that the logarithmically infinite force only arises if the MCL is treated as a two-dimensional manifold. Based on this finite force result, we propose a model for the microscopic dynamic contact angle in \S\ref{sect:DCA_model} and provide supporting experimental comparisons. Similarities between the MCL, Stokeslet, and cusped fluid interface are discussed in \S\ref{sect:force_comp} as they all exhibit singular stresses and finite forces. Concluding remarks are found in \S\ref{sect:conclusion}.


\section{Stokes flow solution to the moving contact line problem}
\label{sect:geo_sol}

\begin{figure}
\centering
\begin{minipage}{0.59\linewidth}\begin{center}
 \includegraphics[width=0.6\linewidth]{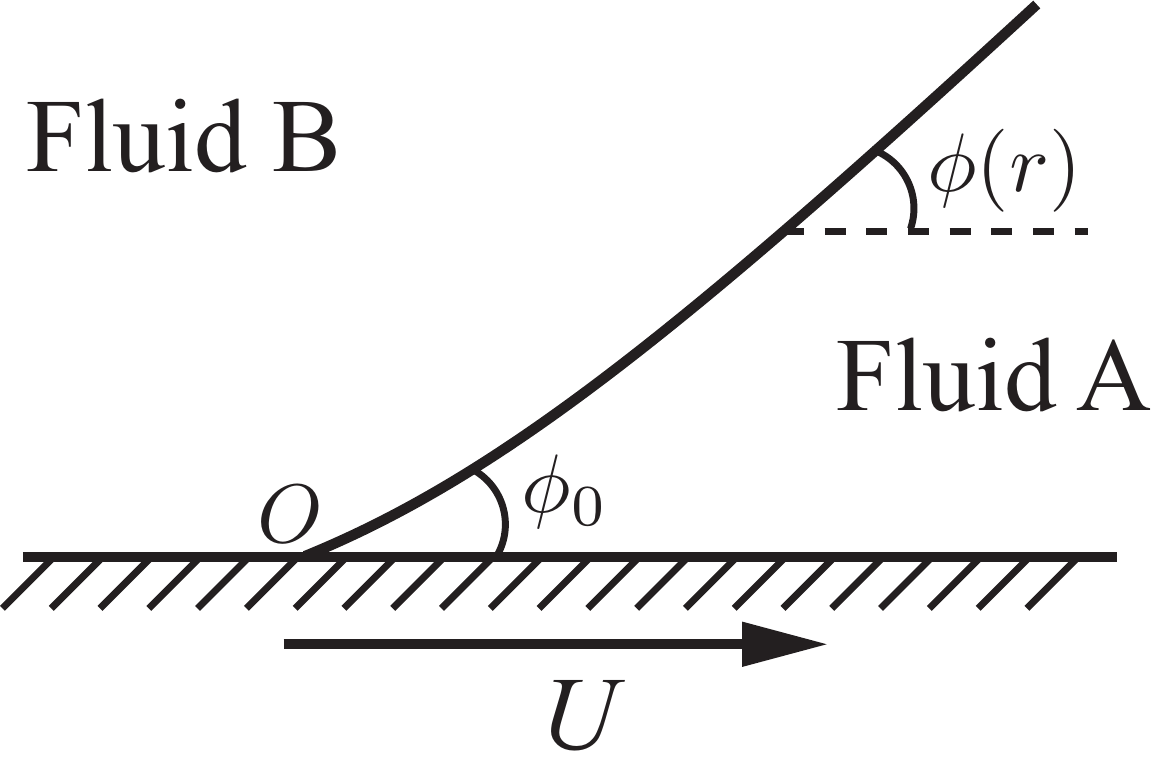}
\end{center}\end{minipage}
\caption{Geometry of a moving contact line in a cylindrical coordinate system ($r,\theta$) whose origin is fixed at the contact line. In this moving reference frame, the solid boundary moves with a velocity $U$ relative to the fluid-fluid interface whose shape is given by $\phi(r)$.}
\label{fig:gen_corner}
\end{figure}

The primary analysis and results of this paper are based off the well-known Stokes solution to the MCL problem originally presented by Cox \cite{CoxRG:86a}. To establish a foundation for the following discussion, this section will provide a brief review of Cox's solution. At a top level, Cox's solution to the MCL flow, shown in figure~\ref{fig:gen_corner}, is obtained using perturbation theory and expanding about the zeroth order solution in the small Capillary number limit. As shown in the past \cite{CoxRG:86a,SnoeijerJ:06a}, the zeroth order flow corresponds to the boundary driven planar wedge whose solution is known analytically from the works of Moffatt \cite{MoffattHK:64a} and Huh \& Scriven \cite{HuhC:71a} and valid in the intermediate and outer regions where the no-slip condition is valid. From this zeroth order solution, Cox and others have iteratively solved for higher order terms. Below, we present Cox's derivation of the zeroth and first order solution.

The analysis performed by Cox begins with assuming that the Reynolds number is significantly smaller than one ($\Rey = \rho \ell U/\mu \ll 1$) so that the dynamics of fluids A and B near the MCL are governed by the Stokes and continuity equations given by
\begin{gather}
\label{eq:stokes1}
\nabla^2 \bm{u}^*_A= \bnabla p^*_A, \hspace{2cm} \bnabla \bcdot \bm{u}^*_A= 0,\\
\label{eq:stokes2}
\lambda \nabla^2 \bm{u}^*_B= \bnabla p^*_B, \hspace{2cm} \bnabla \bcdot \bm{u}^*_B= 0.
\end{gather}
$\rho$ denotes the density, $\ell$ the characteristic length scale, $U$ the interface velocity, $\mu$ the viscosity, $\lambda = \mu_B/\mu_A$ the viscosity ratio, $\bm{u}^*$ the dimensionless velocity, and $p^*$ the dimensionless pressure. If fluid A or B is not specified in the subscript, then the variable is representative of either fluid. In this MCL problem, the fluid is subject to the no-slip and zero penetration boundary condition along all interfaces in addition to the continuity of tangential stress and balance of normal stress at the fluid-fluid interface. In order to make the problem tractable, Cox assumed that Capillary number is significantly smaller than one ($\Ca =\mu U /\sigma \ll 1$) and expanded the velocity, pressure, and fluid-fluid interface shape as
\begin{gather}
\label{eq:u_expan} \bm{u}^* = \bm{u}^*_{0} + \Ca \bm{u}^*_{1}  + \dotsc,\\
\label{eq:p_expan} p^* = p^*_0 + \Ca p^*_1 + \dotsc , \\
\label{eq:phi_expan} \phi = \phi_0 + \Ca \phi_1 + \dotsc .
\end{gather}
Here, $\sigma$ denotes the surface tension. Given the expansions above, other quantities of interest such as the stress tensor, $\bm{T}$, or interface curvature, $\kappa$, can be written using similar expansions. Substituting equations~(\ref{eq:u_expan})-(\ref{eq:p_expan}) into equation~(\ref{eq:stokes1})-(\ref{eq:stokes2}) and collecting terms of order $\Ca^0$ yields the zeroth order governing equations which are given by
\begin{gather}
\nabla^2 \bm{u}^*_{A0}= \bnabla p^*_{A0}, \hspace{2cm} \bnabla \bcdot \bm{u}^*_{A0}= 0,\\
\lambda \nabla^2 \bm{u}^*_{B0}= \bnabla p^*_{B0}, \hspace{2cm} \bnabla \bcdot \bm{u}^*_{B0}= 0.
\end{gather}
As noted by Cox, Capillary number only appears in the normal stress boundary condition at the fluid-fluid interface, i.e.
\begin{gather}
\Ca \bm{\hat{n}} \bcdot [\![ \bm{T}_0 + \Ca \bm{T}_1 + \dotsc  ]\!] \bcdot \bm{\hat{n}'} = \kappa_0 + \Ca \kappa_1 + \dotsc,
\end{gather}
where $[\![ \bcdot ]\!]$ denotes the jump of a quantity across an interface whose normal vector is denoted by $\bm{\hat{n}'}$. Collecting terms of order $\Ca^0$, one obtains $\kappa_0 = 0$ so that the zeroth order solution is the flow in a planar wedge with angle $\phi_0$. To obtain the zeroth order solution, Cox rewrites the Stokes equation as the biharmonic stream function equation, $\nabla^4 \psi = 0$, whose general solution is known and given in \ref{app:bih_sol}. Applying the aforementioned boundary conditions, Cox identifies the zeroth order stream function as
\begin{gather}
\psi_0 =rU[A\cos(\theta) + B \sin(\theta) + C\theta\cos(\theta)+D\theta \sin(\theta)],
\end{gather}
where ($r,\theta$) is the local cylindrical coordinate system.
The coefficients $A$, $B$, $C$, and $D$ are analytically known and presented in \ref{app:coeffs}, in addition to the zeroth order velocity, pressure, and vorticity. The normal stress jump that appears in the zeroth order solution, i.e.
\begin{gather}
-(r/\ell)^{-1}m(\phi_0,\lambda) = -\dfrac{2}{r/\ell} [ \lambda (C_B \cos \phi_0 +D_B \sin \phi_0) - (C_A \cos \phi_0 +D_A \sin \phi_0) ],
\end{gather}
is accounted for by the curvature of the first order interface shape, that is
\begin{gather}
\dfrac{\partial \phi_1}{\partial r} = (r/\ell)^{-1}m(\phi_0,\lambda).
\end{gather}
Integrating the equation above yields $\phi_1 = m(\phi_0,\lambda) \ln (r/\ell) + Q$ and the interface shape
\begin{gather}
\label{eq:int_shape}
\phi = \phi_0+\Ca  [m(\phi_0,\lambda) \ln (r/\ell) + Q]+ \dotsc,
\end{gather}
where $Q$ is an unspecified constant of integration. The equation above is the widely recognized general form of the fluid-fluid interface under steady motion \cite{CoxRG:86a,SibleyD:15a}. In the paper by Cox, higher order terms were not reported as they had a negligible effect. In the next section, we revisit this classic solution to the MCL flow and demonstrate that the singular stress of the dominant zeroth order solution exerts a finite force at the moving contact line.

\section{Forces acting at the contact line}
\label{sect:f_MCL}

\begin{figure}
\centering
\begin{minipage}{0.99\linewidth}\begin{center}
 \includegraphics[width=0.89\linewidth]{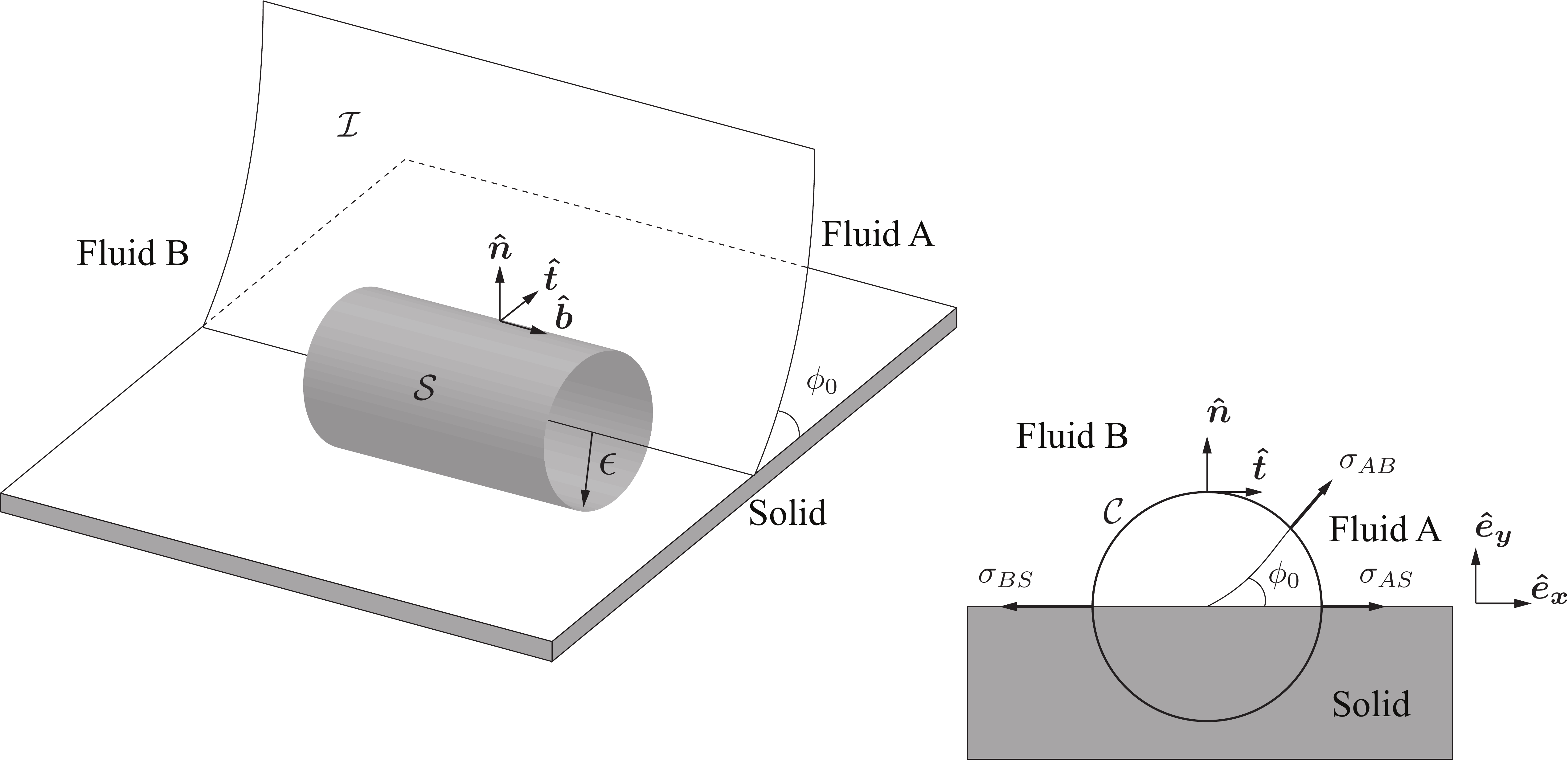}
\end{center}\end{minipage}
\caption{Schematic of the cylindrical control volume $V$ with radius $\epsilon$ that is bounded by the surface $\mathcal{S}$ and centered at the MCL. The moving contact line has a contact angle of $\phi_0$. $\bm{\hat{n}}$ and $\bm{\hat{t}}$ denote the unit normal and tangential vectors of the contour $\mathcal{C}$. $\mathcal{I}$ denotes the interfacial surfaces and $\sigma$ denotes the surface tension force between fluid A, B, and the solid.}
\label{fig:CL_volume}
\end{figure}

In the past, moving contact line analyses have often utilized integral equations that were derived for fluid interfaces. However, the moving contact {\it line} is a one-dimensional manifold unlike fluid interfaces, which are two-dimensional manifolds. A fluid interface is uniquely defined by a single normal vector while a contact line has multiple normal vectors. This multivaluedness subsequently appears in the hydrodynamic solution which exhibits a multivalued velocity, stress, vorticity, and pressure along the contact line. In light of these contact line attributes, let us take a step back and consider a summation of forces acting on a finite sized cylindrical control volume, $V$, with radius $\epsilon$ centered around the contact line, as shown in figure~\ref{fig:CL_volume}. This volume is bounded by the surface $\mathcal{S}$ and intersects the interfacial surfaces denoted by $\mathcal{I}$. As we will demonstrate later, this is the control volume necessary to derive the Young's equation and is the one-dimensional analogue to the rectangular control volume used to derive the balance of forces on a fluid interface which is a two-dimensional manifold. For a steady problem, the sum of all the forces acting on the control volume $V$ is given by
\begin{align}
\label{eq:sum_forces}
\Sigma \bm{f} = \underbrace{\iint_\mathcal{S} \bm{\hat{n}} \bcdot \bm{T} dA}_\text{surface force} +\underbrace{ \iiint_V \rho \bm{g} dV}_\text{body force}   +\underbrace{\iint_\mathcal{I} \bnabla_\pi \bcdot \bm{T}_\pi dA}_{\substack{\text{surface tension} \\ \text{gradient}}}   + \underbrace{ \int_{C\!L} [\![ \bm{T_\pi} \bcdot \bm{\hat{t}'} ]\!] ds}_{\substack{\text{surface tension} \\ \text{force}}} = 0,
\end{align}
assuming massless interfaces \cite{SlatteryJC:07a}. $\bm{g}$ denotes the body force, the subscript $\pi$ denotes surface quantities, $C\!L$ denotes the contact line, and $\bm{\hat{t}'}$ denotes the unit tangential vectors of the interfaces. In the limit as $\epsilon \to \ell_i = 0$, where $\ell_i$ denotes the distance from the contact line over which fluid slip occurs, the equation above represents the sum of all forces acting exactly at the MCL. Note that in some works $\ell_i$ is sometimes denoted by the slip length, $\ell_s$, because they are often considered to be the same order of magnitude. However they are not equivalent and $\ell_i$ and $\ell_s$ are treated as separate length scales in this work. For this analysis, we will assume that there are zero body forces ($\bm{g} = 0$) and zero surface tension gradients ($\bnabla_\pi \bcdot \bm{T}_\pi = 0$). As a result, the $x$ and $y$ component of the forces (per unit contact line length) acting on the control volume are reduced to
\begin{gather}
\label{eq:MCL_fx}
\Sigma f_x = \lim_{\epsilon \to 0}\oint_\mathcal{C} \bm{\hat{n}}\bcdot \bm{T} \bcdot \bm{\hat{e}_x}  ds +  \sigma_{AS} - \sigma_{BS} + \sigma_{AB} \cos(\phi_0) = 0,\\
\label{eq:MCL_fy}
\Sigma f_y = \lim_{\epsilon \to 0}\oint_\mathcal{C} \bm{\hat{n}}\bcdot \bm{T} \bcdot \bm{\hat{e}_y} ds + \sigma_{AB} \sin(\phi_0) = 0,
\end{gather}
where $\mathcal{C}$ denotes the contour path. The effects of surface tension gradients are considered in a related work by Thalakkottor \& Mohseni \cite{mohseni:17r}.

For the forces acting in the $x$ direction, we decompose the stress integral into three segments that lie inside each material so that the integral above is rewritten as
\begin{align} 
\notag
\Sigma f_x = &\int_0^{\phi_0} \bm{\hat{e}_r} \bcdot (\bm{T}_{A0} + \Ca \bm{T}_{A1} + \dotsc) \bcdot \bm{\hat{e}_x} r d\theta + \int_{\phi_0}^\pi  \bm{\hat{e}_r} \bcdot (\bm{T}_{B0} + \Ca \bm{T}_{B1} + \dotsc) \bcdot \bm{\hat{e}_x} r d\theta \\ 
\label{eq:fx_split} 
  &+\int_\pi^{2\pi}  \bm{\hat{e}_r} \bcdot \bm{T}_S \bcdot \bm{\hat{e}_x} r d\theta+ \sigma_{AS} - \sigma_{BS} + \sigma_{AB}\cos(\phi_0) = 0.
\end{align}
Here, the integral of $\bm{T}_S$ represents the force induced by the stress of the enclosed solid, $\bm{f}_\text{S}$. To evaluate the fluid stress tensor integrals, we begin with the zeroth order solution and use the stress tensor decomposition $\bm{T} = \bm{\hat{T}} -2\mu \bm{B}$, where $\bm{\hat{T}} = -p\bm{I} + 2\mu\bm{\Omega}$ is the reduced stress tensor, $\bm{\Omega}$ is the vorticity tensor, and $\bm{B} = (\nabla \bcdot \bm{u})\bm{I} - (\nabla\bm{u})^T$ is the surface deformation rate tensor. With this decomposition, the first stress tensor integral in equation~(\ref{eq:fx_split}) is rewritten as
\begin{gather}
\label{eq:stress_decomp}
\int_0^{\phi_0} \bm{\hat{e}_r} \bcdot \bm{T}_{A0} \bcdot \bm{\hat{e}_x} r d\theta = \int_0^{\phi_0} \bm{\hat{e}_r} \bcdot \bm{\hat{T}}_{A0} \bcdot \bm{\hat{e}_x} r d\theta - 2\mu_A\int_0^{\phi_0} \bm{\hat{e}_r} \bcdot \bm{B}_{A0} \bcdot \bm{\hat{e}_x} r d\theta,
\end{gather}
for the zeroth order solution. The force contribution of the surface deformation rate tensor is identically zero, as one can show that
\begin{gather*}
\bm{\hat{e}_r} \bcdot \bm{B}_{A0} = -\dfrac{\partial u_{0r}}{\partial r} \bm{\hat{e}_r} - \dfrac{\partial u_{0\theta}}{\partial r} \bm{\hat{e}_\theta} = 0,
\end{gather*}
since both the radial and azimuthal components of the zeroth order velocity are independent of $r$. This is consistent with the findings of Wu et al \cite{WuJZ:06a}, who reported that the surface deformation rate tensor does not contribute to the total surface force over a closed boundary if viscosity is constant. By substituting the zeroth order solution for pressure and vorticity into the first term on the right hand side of equation~(\ref{eq:stress_decomp}), we obtain
\begin{align}
\notag
f_{A0,x} &= \int_0^{\phi_0} \bm{\hat{e}_r} \bcdot \bm{\hat{T}}_{A0} \bcdot \bm{\hat{e}_x} r d\theta = \int_0^{\phi_0}  (-p_{A0}\bm{\hat{e}_r} +\mu_A\omega_{A0} \bm{\hat{e}_\theta}) \bcdot \bm{\hat{e}_x} r d\theta\\ \notag
&=-\mu_A U\int_0^{\phi_0} \left\{ \dfrac{2}{r} [C_A\cos(\theta) + D_A\sin(\theta)]\cos(\theta) + \dfrac{2}{r} [C_A\sin(\theta)-D_A\cos(\theta)]\sin(\theta) \right\} r d\theta\\ \notag
&= -2\phi_0 \mu_A U C_A.
\end{align}

The result above shows that the zeroth order viscous force contribution from fluid A in the $x$ direction ($f_{A0,x}$) is independent of $r$ and a function of contact angle, viscosity, and interface velocity only. The same analysis performed on the segment that lays in fluid B and for the zeroth order fluid stress integrals in the $y$ direction allows us to rewrite equations~(\ref{eq:MCL_fx}) and (\ref{eq:MCL_fy}) as
\begin{gather}
\label{eq:fx_final}
2\phi_0 \mu_A U C_A + 2(\pi-\phi_0)\mu_B U C_B - \sigma_{AB}\cos(\phi_0) = \sigma_{AS} - \sigma_{BS} + f_{\text{S},x},\\
\label{eq:fy_final}
2\phi_0 \mu_A U D_A + 2(\pi-\phi_0)\mu_B U D_B -\sigma_{AB}\sin(\phi_0) = f_{\text{S},y},
\end{gather}
and represents the balance of forces in the $x$ and $y$ direction when the sum of all forces is zero and the moving contact line is steady. The equations above are zeroth order accurate and have been organized so that the left hand side contains the forces that the fluid exerts on the solid, and the right hand side contains the forces that the solid exerts on the fluid. Note that the expressions above are independent of $r$, so that even in the limit as $\epsilon \to 0$, the total MCL force of the zeroth order solution remains {\it finite} despite a singular stress. Conceptually, one can think of this as an infinite stress acting on an infinitesimally small area, resulting in a finite force. 

The analysis above, which has been performed for the zeroth order solution, can also be applied to the first order solution. Given the first order interface shape in equation~(\ref{eq:int_shape}), and assuming that $\bm{u}_1$ is finite at the contact line, the first order stream function will have the form given by
\begin{gather}
\psi_{1} = (r/\ell)^{2} \ln (r/\ell) q_{2L,1}(\theta)+ \sum_{n = 1}^\infty (r/\ell)^{n} q_{n,1}(\theta).
\end{gather}
From this stream function, the velocity field and stress tensor can be analytically determined so that first order equivalent of equation~(\ref{eq:stress_decomp}), taken in the limit as $r \to 0$, becomes
\begin{gather}
\label{eq:stress_first}
\lim_{r \to 0} \int_0^{\phi_0} \bm{\hat{e}_r} \bcdot (\Ca\bm{T}_{A1}) \bcdot \bm{\hat{e}_x} r d\theta =  \lim_{r \to 0} \int_0^{\phi_0} \Ca \left[\ln(r/\ell) h_{2L}(\theta)+ \sum_{n = 1}^\infty (r/\ell)^{n-2} h_n(\theta) \right] r d\theta.
\end{gather}
Here, $h$ simply denotes the function that collects the terms of $\bm{\hat{e}_r} \bcdot (\Ca\bm{T}_{A1}) \bcdot \bm{\hat{e}_x}$ that scale with the same order of $r$. In this limit where $r \to 0$, all terms of the integrand vanish except the $n = 1$ term. Thus the first order viscous force of fluid A acting at the MCL is given by 
\begin{gather}
\Ca f_{A1,x} = \Ca \int_0^{\phi_0} h_1(\theta) d\theta.
\end{gather}
As $f_{A0,x}$ and $f_{A1,x}$ are of the same order of magnitude and because $\Ca \ll 1$, we conclude that the first order force is significantly smaller than the zeroth order force. Similarly, we find that the first order forces due to fluid B and in the $y$ direction are also negligible so that the balance of forces at the MCL is given by equation~(\ref{eq:fx_final}) and (\ref{eq:fy_final}). In the work by Cox \cite{CoxRG:86a}, the first order velocity field and corresponding viscous forces are also neglected as they have a negligible effect on the interface shape when compared to the zeroth order forces.

Based on the finite force results above, we remark that the zeroth order MCL solution is a prime example of the singular mathematical models described by Dussan \& Davis \cite{Dussan:74a}, much like the well-known Stokeslet \cite{Guazzelli:11a}. In both the Stokeslet and the zeroth order MCL solution, the force acting over a small area is modeled as mathematical line with infinite stress and finite force. Therefore, they can be considered physically realistic, at least in regard to conservation of mass and momentum. In the context of moving contact lines, it may appear unusual for a model to contain a singular stress and finite force. However, we emphasize that the Stokeslet, potential line source, and potential line vortex all exhibit the same characteristics and have all been successfully used in modeling a wide variety of physical phenomena. In the following discussion, we present a complex formulation of the MCL problem that yields the same finite force at the MCL, but avoids some of the algebra through the use of Cauchy's residue theorem.

\subsection*{Complex formulation of the moving contact line force}

In the analysis above, the steps required to find the MCL force can be some what cumbersome and therefore, we introduce a relatively simpler complex formulation of the problem in this section.  The advantages of this formulation will become clear in \S\ref{sect:force_comp}, when the MCL problem is compared with the Stokeslet and cusped fluid interface.

\begin{figure}
\centering
\begin{minipage}{0.99\linewidth}\begin{center}
 \includegraphics[width=0.99\linewidth]{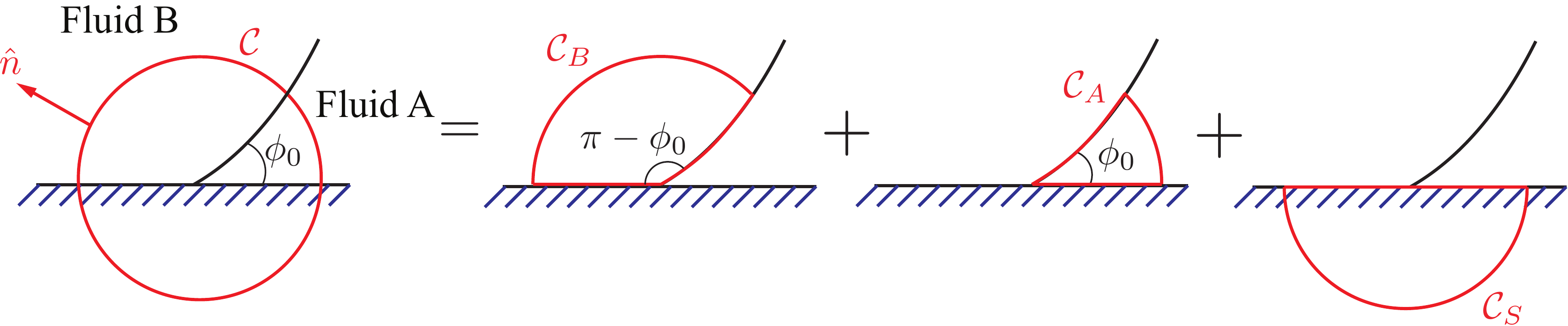}
\end{center}\end{minipage}
\caption{Decomposition of the contour $\mathcal{C}$ into $\mathcal{C}_A$, $\mathcal{C}_B$, and $\mathcal{C}_S$.}
\label{fig:contour_decomp}
\end{figure}

As demonstrated by Langlois \& Deville  \cite{LangloisWE:64a}, any flow satisfying the Stokes equation will ensure that the pressure and vorticity are harmonic conjugates. Thus, we can define the function  $G = \mu \omega + ip$ representing the shear and normal stresses. For the zeroth order solution, $G_0$ is given by
\begin{gather}
\label{eq:G}
G_0 = \mu \omega_0 + ip_0 =  2\mu U\dfrac{-D+iC}{z}.
\end{gather}
It is immediately apparent that $G_0$ has a simple pole at the location of the contact line where $z = 0$ and represents a dipole distribution. Furthermore, we observe that the complex function $G$ and the reduced stress tensor $\bm{\hat{T}}$ are composed of pressure and vorticity only. Thus, it is not all that surprising that the contour integral of $\bm{\hat{T}}$ can be written in terms of a complex contour integral of $G$, that is
\begin{align}
\notag\oint G dz &= \oint ( \mu\omega + ip)(dx + i dy)\\
\notag &= \oint (-p\boldsymbol{\hat{e}_x}- \mu\omega \boldsymbol{\hat{e}_y}) \bcdot \left(\dfrac{dy}{ds} \boldsymbol{\hat{e}_x}- \dfrac{dx}{ds} \boldsymbol{\hat{e}_y} \right) ds +  i\oint ( \mu\omega \boldsymbol{\hat{e}_x} -p \boldsymbol{\hat{e}_y}) \bcdot \left(\dfrac{dy}{ds} \boldsymbol{\hat{e}_x}- \dfrac{dx}{ds} \boldsymbol{\hat{e}_y} \right) ds\\
\label{eq:f_stress} & = {\oint \boldsymbol{\hat{n}}  \bcdot  \mathbf{\hat{T}} \bcdot \boldsymbol{ \hat{e}_x} ds} + i{\oint \boldsymbol{\hat{n}}  \bcdot  \mathbf{\hat{T}} \bcdot \boldsymbol{ \hat{e}_y} ds}.
\end{align}
Here, the real and imaginary components are exactly equal to the stress integral terms of equations~(\ref{eq:MCL_fx}) and (\ref{eq:MCL_fy}) and correspond to the viscous force exerted by the fluid in the $x$ and $y$ directions. In contrast to the previous analysis, this formulation reveals that we can avoid the tedious algebra of integrating $\bm{\hat{T}}$, and instead evaluate the left hand side of equation~(\ref{eq:f_stress}) using Cauchy's method of residues \cite{MitrinovicD:84a}. However, the standard residue theorem cannot be applied over the contour $\mathcal{C}$, as the function $G$ is piecewise holormophic within the contour. Therefore, it is necessary to decompose the contour $\mathcal{C}$ into the three subcontours $\mathcal{C}_A$, $\mathcal{C}_B$, and $\mathcal{C}_S$ that enclose each phase, see figure~\ref{fig:contour_decomp}. Inside each subcontour, $G$ is entirely holomorphic and the pole resides at the MCL ($z = 0$). Thus, we can treat the contour integral of $G_{A}$ and $G_{B}$ using Cauchy's residue theorem and the Sokhotski-Plemelj formulas \cite{EstradaR:12a}. For the zeroth order solution, we obtain
\begin{gather*}
\oint_\mathcal{C} G_0 dz =\oint_{\mathcal{C}_A} G_{A0} dz + \oint_{\mathcal{C}_B} G_{B0} dz + \oint_{\mathcal{C}_S} G_S dz,\\
\oint_{\mathcal{C}_A} G_{A0} dz = \phi_0 i \Res(G_{A0}) = -2\phi_0 \mu_A U C_A -i2\phi_0 \mu_A U D_A,\\
\oint_{\mathcal{C}_B} G_{B0} dz = (\pi-\phi_0) i \Res(G_{B0}) = -2(\pi-\phi_0) \mu_B U C_B -i2(\pi-\phi_0) \mu_B U D_B.
\end{gather*}
Additional details regarding the Sokhotski-Plemelj formulas and the treatment of singularities residing on the contour can be found in \ref{app:SP_theorem}. As before, the integral of $G_S$ is the force exerted by the stress of the solid.

The real component of the complex contour integrals above can be substituted into equation~(\ref{eq:fx_split}), and we once again obtain equation~(\ref{eq:fx_final}), representing the balance of viscous and surface tension forces at the MCL. Similarly the imaginary components can be used to obtain equation~(\ref{eq:fy_final}). Interestingly, one can use the complex contour integral of $G$ to find the total surface force for any Stokes flow. As we will show in \S\ref{sect:force_comp}, this complex analysis correctly captures the viscous force exerted by two other singular Stokes flows that have similar pressure and vorticity fields. While we have only presented a relatively simple complex formulation relevant to the MCL force, complex variables can also be used to define the stream function and velocity to analytically solve for viscous flows in a variety of problems \cite{CrowdyDG:10a,CrowdyD:17a}. In the following section, we compare our result to previous works and discuss the physical implications and limitations of this solution.

\section{Discussion of the MCL force}
\label{sect:discussion}

The analysis of the previous section shows that the force at the moving contact line predicted by the Stokes solution is finite. However, previous works \cite{Batchelor:67a,HuhC:71a} have reported a logarithmically infinite total force on the solid. So why does the analysis above predict a finite force when others report an infinite force? To understand the distinction, we first replicate the result of Huh \& Scriven, by integrating the stress of fluid A along {\it only} the fluid-solid boundary, i.e.
\begin{gather}
\label{eq:surf_f_int}
f_{AS,x} = \int_0^R \bm{\hat{e}_\theta} \bcdot \bm{T}_{A0} \bcdot \bm{\hat{e}_x} dr \quad \text{at} \quad  \theta = 0^\circ,
\end{gather}
where $R$ is some finite length. From the solution provided in \ref{app:coeffs}, one can show that $\bm{\hat{e}_\theta} \bcdot \bm{T}_{A0} \bcdot \bm{\hat{e}_x} = -\mu_A \omega_{A0}$, and that the integral above is improper, as $\omega_{A0}$ is singular at $r = 0$. Therefore, this integral can only be evaluated in the limit, that is
\begin{align*}
f_{AS,x} &=  \lim_{\epsilon \to 0}\int_\epsilon^R -\dfrac{2\mu_A U}{r} [C_A\sin(0) - D_A\cos(0)] dr\\
 &= \lim_{\epsilon \to 0} 2\mu_A D_A U[\ln R - \ln \epsilon]  \\
 & =  \infty.
 \end{align*}
Thus, the viscous force exerted by fluid A along the fluid-solid interface is logarithmically infinite. The same analysis performed for fluid B at $\theta = 180^\circ$ yields another infinite force. {\it Individually}, fluid A and B exert infinite forces along the fluid-solid interface, however, the sum of these two infinite forces is undefined or infinite depending on the sign of $D$. Note that in this approach, only the forces along the fluid-solid interface are considered. This would contradict Young's equation, which clearly includes the surface tension force of the fluid-fluid interface that only exists at $r = 0$ on the solid boundary, and nowhere else.

Based on the discussion above, we find that the distinction between our analysis and previous results, is the control volume that is used to derive the force integral. In the analysis of Huh \& Scriven, the force integral given by equation~(\ref{eq:surf_f_int}) is derived for rectangular control volumes containing discontinuities across a two-dimensional manifold or surface, i.e. a fluid interface \cite{LealLG:07a}. In contrast, the force integral presented in equation~(\ref{eq:stress_decomp}) is specifically derived for volumes containing line discontinuities like the contact line and thus includes the forces acting along the fluid-fluid interface. In works that report an infinite force, it appears that the MCL was viewed as an extension of the interface between a single fluid and solid. Thus, the force was determined by integrating the fluid stress along the fluid-solid interface only. However, the MCL is {\it not} an extension of a fluid-solid interface, but rather a line defined by the intersection of three immiscible materials. From this perspective, it is natural to define a control volume that encloses all three materials and the MCL. In related analyses of the contact line Slattery et al \cite{SlatteryJC:07a} and Andreotti \& Snoeijer \cite{SnoeijerJH:16a} have also chosen the same cylindrical control volume. By defining the control volume in this fashion, we include the forces that act on the fluid-fluid interface and capture the multivalued nature of the MCL.

\begin{figure}
\centering
\begin{minipage}{0.89\linewidth}\begin{center}
 \includegraphics[width=0.49\linewidth]{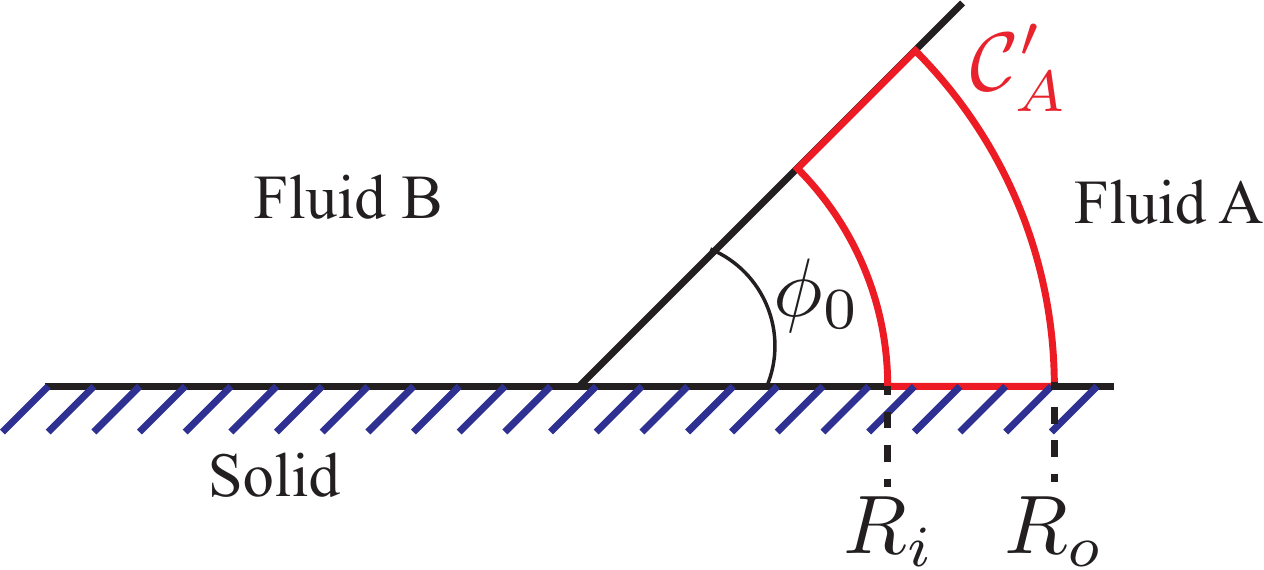}
\end{center}\end{minipage}
\caption{Annular contour $\mathcal{C}'_A$ (shown in red) with inner radius $R_i$ and outer radius $R_o$ residing in fluid A.  }
\label{fig:annul_cont}
\end{figure}

To physically understand why the total force at the MCL remains finite, consider the annular contour $\mathcal{C}'_A$ with inner radius $R_i$ and outer radius $R_o$ that lies inside fluid A, as shown in figure~\ref{fig:annul_cont}. In the zeroth order solution, the stress along the two radial segments and the stress along the two azimuthal segments are exactly equal and opposite, that is
\begin{gather*}
\underbrace{-\left[\bm{\hat{e}_\theta} \bcdot \bm{\hat{T}}_{A0} \bcdot \bm{\hat{e}_x} \right]_{\theta = 0}}_\text{fluid-solid interface} =  \underbrace{\left[\bm{\hat{e}_\theta}  \bcdot \bm{\hat{T}}_{A0} \bcdot \bm{\hat{e}_x} \right]_{\theta = \phi_0}}_\text{fluid-fluid interface}  = -\dfrac{2 \mu_A UD_A}{r},\\
\underbrace{-\left[(\bm{\hat{e}_r}  \bcdot \bm{\hat{T}}_{A0} \bcdot \bm{\hat{e}_x})r \right]_{r = R_i}}_\text{inner arc} =  \underbrace{\left[(\bm{\hat{e}_r} \bcdot \bm{\hat{T}}_{A0} \bcdot \bm{\hat{e}_x})r \right]_{r = R_o}}_\text{outer arc} = -2 \mu_A UC_A.
\end{gather*}
Thus, the force acting along the fluid-fluid interface is balanced by the force acting along the fluid-solid interface. Similarly, the forces along the two azimuthal arcs balance each other and the total surface force acting on the contour $\mathcal{C}'_A$ is zero. In fact, any contour path that does not enclose or pass through the singularity will yield zero total surface force. If we shrink the radius of the inner arc to zero, the contour $\mathcal{C}'_A$ is reduced to $\mathcal{C}_A$ and passes through the singularity, as shown in figure~\ref{fig:contour_decomp}. Here, the logarithmically infinite force along the fluid-solid interface is balanced by the force along the fluid-fluid interface, as the stress along each boundary approaches positive and negative infinity at the same rate. A similar cancellation of logarithmic singularities is reported in the work of Jones \cite{JonesMA:03a} when the free vortex sheet is shed tangentially from the plate edge. In the end, the total force exerted by the fluid is given by the azimuthal arc and remains finite no matter how small $\mathcal{C}_A$ becomes. 

While the analysis above demonstrates that there is a finite force at the MCL, the model is not without its limitations. One such limitation is the viscous dissipation per unit volume, which scales as $r^{-2}$. This dissipation is non-integrable for both fluids and results in a singular total energy. However, this does not affect the balance of momentum so long as density and viscosity are constant. This singular energy is not unique to the MCL problem and appears in several other two-dimensional singular continuum models. For example, the potential line source/sink and line vortex have infinite kinetic energy at the singularity \cite{Batchelor:67a}. From electromagnetism, the energy per unit volume for an infinite line charge or infinitely long current carrying wire is also singular \cite{GriffithsDJ:72a}. These singular magnitudes in energy are a consequence of modeling some finite sized physical feature as a mathematical line, where some desired integral quantity is preserved. For example, the potential line vortex is the limiting case of a Rankine vortex where the radius of the central core is reduced to zero while preserving the total circulation. In this limit, circulation can only be preserved if angular velocity approaches infinity, therefore, the potential line vortex exhibits infinite kinetic energy. Despite this non-physical kinetic energy, potential flow theory has successfully modeled a wide range of high Reynolds number flows. Similarly, the MCL model presented in this manuscript is the limit where the fluid slip region has been reduced to an infinitely small point while preserving the total force. We recognize that these singular continuum models are idealized representations of the true physical phenomena. For the MCL problem, this relatively simple model will require additional development in order to capture the transfer of energy. However, it is still valid when considering forces and momentum transfer near the MCL. In the following section, we explore the impact of this model on the prediction of dynamic contact angle.

\section{Dynamic contact angle model}
\label{sect:DCA_model}

The force balance presented above is essentially a dynamic Young's equation that can be used to model the dynamic contact angle. To do so, we assume that the solid is relatively rigid such that any deformation of the solid near the MCL is extremely small \cite{SlatteryJC:07a,LesterG:61a}. As a result, ${f_{\text{S},x}}$ is small relative to the other terms of equation~(\ref{eq:fx_final}) and the force balance at the contact line can be written as
\begin{gather}
\label{eq:DCA}
2\phi_0 \mu_A U C_A + 2(\pi-\phi_0)\mu_B U C_B - \sigma_{AB}\cos(\phi_0) = \sigma_{AS} - \sigma_{BS}.
\end{gather}
In the equation above, the viscosity, interface velocity, and surface tension are known and therefore one can solve for the only remaining unknown variable, namely the dynamic contact angle $\phi_0$. This angle corresponds to the microscopic contact angle since equation~(\ref{eq:DCA}) represents the force balance at $r = \ell_i$. For the idealized hydrodynamic solution, $\ell_i = 0$ and $\phi_0$ is the microscopic angle measured at the solid surface. In problems where slip occurs over a finite but significantly smaller length than the capillary length ($\ell_i \ll \ell_c$), $\phi_0$ corresponds to the microscopic angle measured just outside the slip region. In the limit $U \to 0$, equation~(\ref{eq:DCA}) is simplified to the static Young's equation. The static Young's equation can be used to replace the right hand side of equation~(\ref{eq:DCA}) with $-\sigma_{AB}\cos(\phi_\text{static})$ so that the dimensionless dynamic Young's equation is rewritten as
\begin{gather}
\label{eq:nond_DCA}
\cos(\phi_0) - \cos(\phi_\text{static}) = \Ca_A [2\phi_0 C_A + 2(\pi-\phi_0)C_B\lambda].
\end{gather}

This non-dimensional form reveals that the change in microscopic contact angle scales with Capillary number and is consistent with diffuse interface and molecular kinetic models. In its simplest form, the diffuse interface model predicts that microscopic contact angle will scale as
\begin{gather}
\label{eq:DI_DCA}
\cos(\phi_0) - \cos(\phi_\text{static}) \sim \Ca \dfrac{\zeta}{\ell_i},
\end{gather} 
where $\zeta$ is the interface width \cite{EW:07a,SnoeijerJH:13b}. Molecular kinetic theory proposes the relation
\begin{gather}
\label{eq:MKT_DCA}
\cos(\phi_0) - \cos(\phi_\text{static})  = \Ca F_B(\nu_0,\xi),
\end{gather}
where $F_B$ is a Boltzman factor that is a function of the molecular equilibrium frequency and molecular displacement, $\nu_0$ and $\xi$ \cite{BlakeTD:06a,BlakeTD:11a,SnoeijerJH:13b}. The right hand side of equations~(\ref{eq:nond_DCA}-\ref{eq:MKT_DCA}) show that the change in microscopic contact angle scales with Capillary number multiplied by a factor representing the proposed physical mechanisms of each model. In our proposed model, the factor on the right hand side represents the total viscous force acting on the contact line region and only contains macroscopic parameters that are known a priori, e.g. viscosity. In contrast, microscopic parameters such as the interface thickness, molecular displacement, and molecular equilibrium frequency need to be empirically determined from experimental data. Interestingly, we have not specified the specific physical mechanism which regularizes the stress singularity at microscopic scales, whether it be slip or molecular attachment and detachment. Our analysis essentially finds that regardless of the microscopic physical mechanisms, the total force exerted must equal the net change in fluid momentum. As conservation of momentum should be satisfied for any physical model, it is somewhat expected that equations~(\ref{eq:nond_DCA}-\ref{eq:MKT_DCA}) are similar in form. Note that this perspective on the MCL is conceptually similar to vortex sheets which are used to model boundary layers in potential flow theory. While the exact velocity profile within the boundary layer is not resolved by the vortex sheet, the total integrated circulation is captured and potential flow theory is able to accurately model the forces acting on solid bodies.

To demonstrate the utility of this model, we apply it to the Brookfield std. viscosity fluid and 70\% glycerol solution, whose dynamic contact angles were experimentally measured by Hoffman \cite{HoffmanRL:75a} and Blake \& Shikhmurzaev \cite{ShikhmurzaevYD:02a}, respectively. In this example, these two fluids and experiments are chosen for their contrasting fluid properties and experimental set ups. However, the same analysis can be applied to other fluids and geometries as well. As shown in table~\ref{tab:fluid_prop}, the Brookfield fluid has a high viscosity and perfectly wets the solid while the glycerol solution has a significantly lower viscosity and only partially wets the solid. With respect to the experimental set up, Hoffman measured the contact angle of a liquid slug as it was pushed through a 1.95 mm precision bore tube while Blake \& Shikhmurzaev measured the dynamic contact angle created by plunging a smooth tape into a bath of fluid. Despite these two very different fluids and experiments, we will see that the theoretical model is in good agreement with the experimental measurements.

\begin{figure}
\centering
\begin{minipage}{0.49\linewidth}\begin{center}
 \includegraphics[width=0.99\linewidth]{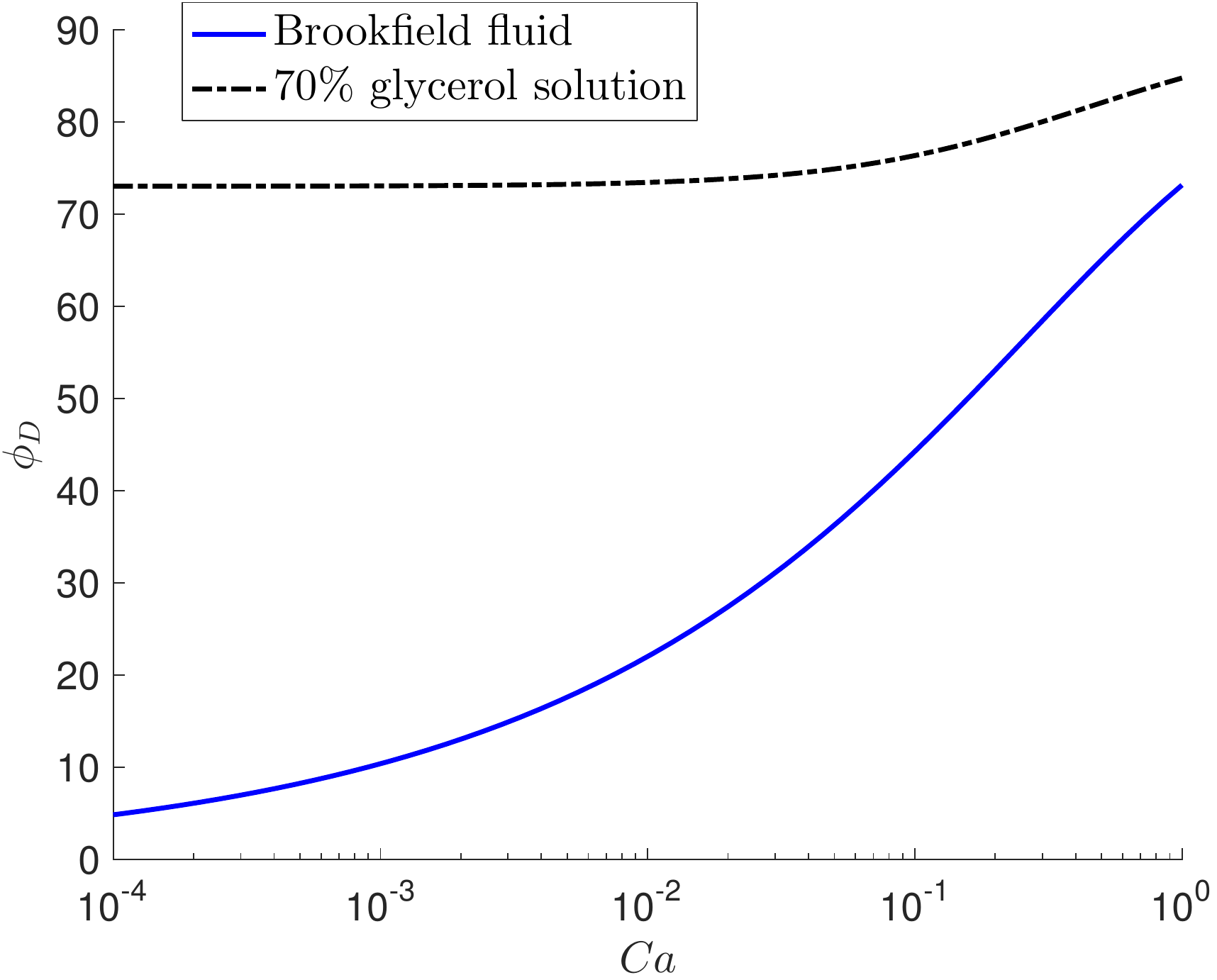}
\end{center}\end{minipage}
\begin{minipage}{0.49\linewidth}\begin{center}
 \includegraphics[width=0.99\linewidth]{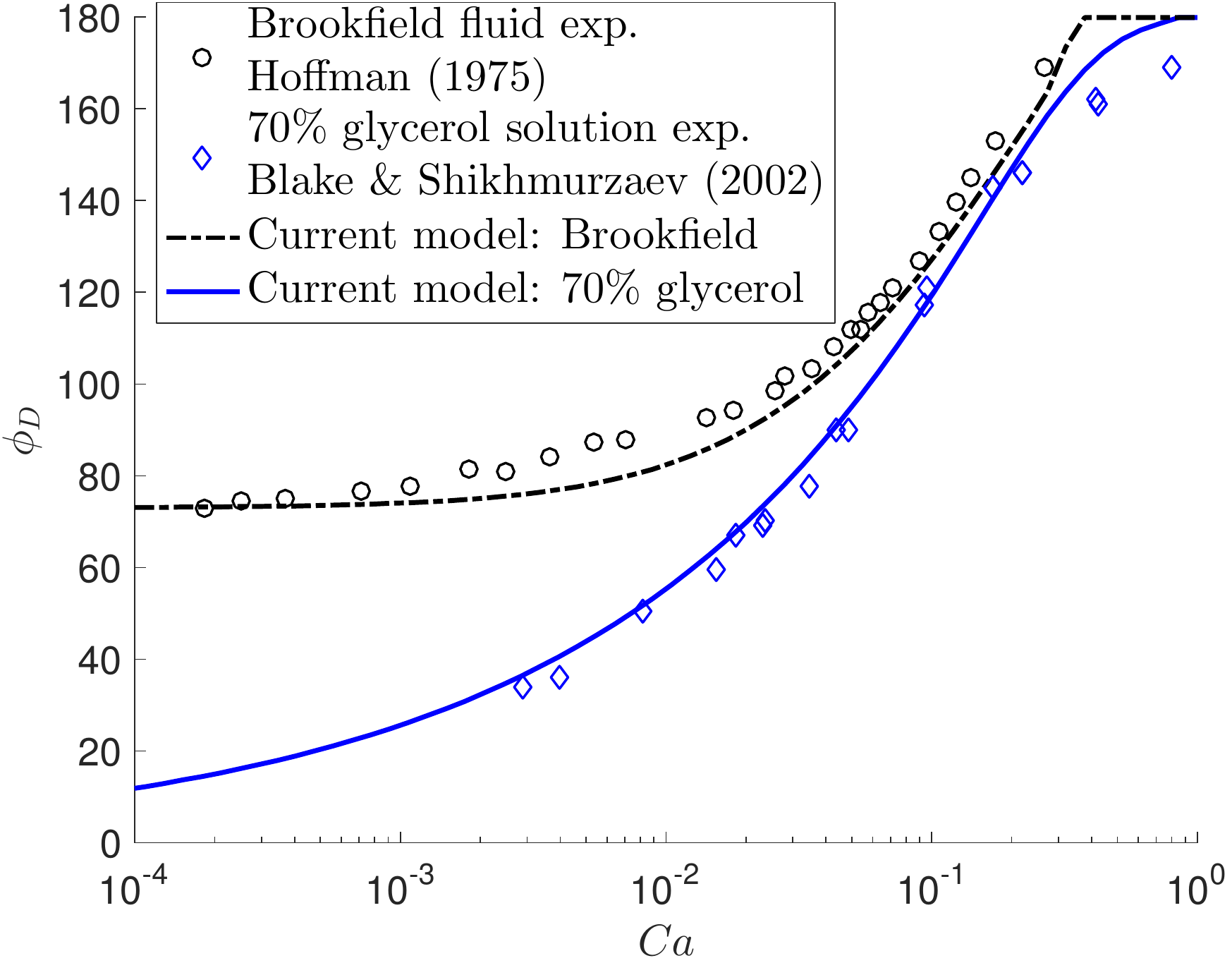}
\end{center}\end{minipage}
\begin{minipage}{0.49\linewidth}\begin{center}
(a)
\end{center}\end{minipage}
\begin{minipage}{0.49\linewidth}\begin{center}
(b)
\end{center}\end{minipage}
\caption{(a) Microscopic dynamic contact angle predicted for the Brookfield fluid and 70\% glycerol solution investigated by Hoffman \cite{HoffmanRL:75a} and Blake \& Shikhmurzaev \cite{ShikhmurzaevYD:02a}, respectively. Microscopic dynamic contact angle is obtained using equation~(\ref{eq:DCA}) and the MCL solution presented in \S\ref{sect:geo_sol}. (b) Apparent dynamic contact angle comparison between experimental measurements and current theoretical model. The current model uses Cox's model \cite{CoxRG:86a} where the microscopic contact angle, $\phi_0$, is theoretically predicted by equation~(\ref{eq:DCA}). }
\label{fig:dynamic_CA}
\end{figure}

Given the fluid properties in table~\ref{tab:fluid_prop}, we use equation~(\ref{eq:DCA}) to obtain a theoretical prediction of microscopic dynamic contact angle as a function of Capillary number, as shown in figure~\ref{fig:dynamic_CA}(a). To account for the roughness of the solid surface we have used $\phi(\Ca \to 0^+)$ in place of $\phi_\text{static}$ in equation~(\ref{eq:DCA}). At first glance, we observe that the glycerol solution exhibits a relatively smaller change in contact angle as $\Ca$ increases. This is due to the significantly larger viscosity ratio that allows the Brookfield fluid to more easily reduce the contact angle of the receding air phase. To compare our results with those reported by Hoffman and Blake \& Shikhmurzaev, we combine our model for microscopic contact angle ($\phi_0$) with Cox's model for apparent contact angle ($\phi_D$) \cite{CoxRG:86a}. Cox's zeroth order model is given
\begin{gather}
\label{eq:cox_g}
g(\lambda,\phi_D) = g(\lambda,\phi_0) + \Ca \ln (1/\varepsilon),
\end{gather}
where all terms are of order 1 and $\varepsilon = \ell_i/\ell_o$ is the ratio of the inner length scale to the outer length scale. $g(\lambda,\phi)$ is given by
\begin{gather*}
g(\lambda,\phi) = \int_0^\phi \dfrac{\lambda(\beta^2-\sin^2\beta)[(\pi-\beta)+\sin\beta\cos\beta]+[(\pi-\beta)^2-\sin^2\beta](\beta-\sin\beta\cos\beta)}{2\sin\beta[\lambda^2(\beta^2\sin^2\beta)+2\lambda\{\beta(\pi-\beta)+\sin^2\beta\}+\{ (\pi-\beta)^2-\sin^2\beta\}]} d\beta.
\end{gather*}
The apparent contact angle predicted by the combination of these two models is shown in figure~\ref{fig:dynamic_CA}(b) with $\varepsilon = 10^{-4}$. Note that the outer length scale is the distance at which the apparent contact angle is measured and is often interpreted as the capillary length scale. Hoffman and Blake \& Shikhmurzaev did not report the length scale of their contact angle measurements, however it was likely smaller than the capillary length scale due to their use of microscopes. Consequently, experimentally obtained values of $\varepsilon$ are slightly larger than one might expect if one were to use the capillary length as the outer length scale.

\begin{table}\centering 
\begin{tabular}{@{} lcccc @{}}
& $\mu$ [N s/m$^2]$ \hspace{1cm}& $\rho$ [kg/m$^3]$ \hspace{1cm}& $\sigma_{AB}$ [N/m] \hspace{1cm}& $\phi_\text{static}$ \hspace{1cm}\\ \hline
Brookfield std. viscosity fluid \quad \quad& 98.8 & 974 & 0.0217 & $0^\circ$\\
70\% glycerol solution  & 0.023 & 1181 & 0.0635 & $67^\circ$\\
\end{tabular}
\caption{Fluid properties of Brookfield std. viscosity fluid and 70\% glycerol solution investigated by Hoffman \cite{HoffmanRL:75a} and Blake \& Shikhmurzaev \cite{ShikhmurzaevYD:02a}, respectively.}
\label{tab:fluid_prop}
\end{table}

\begin{figure}
\centering
\begin{minipage}{0.49\linewidth}\begin{center}
 \includegraphics[width=0.99\linewidth]{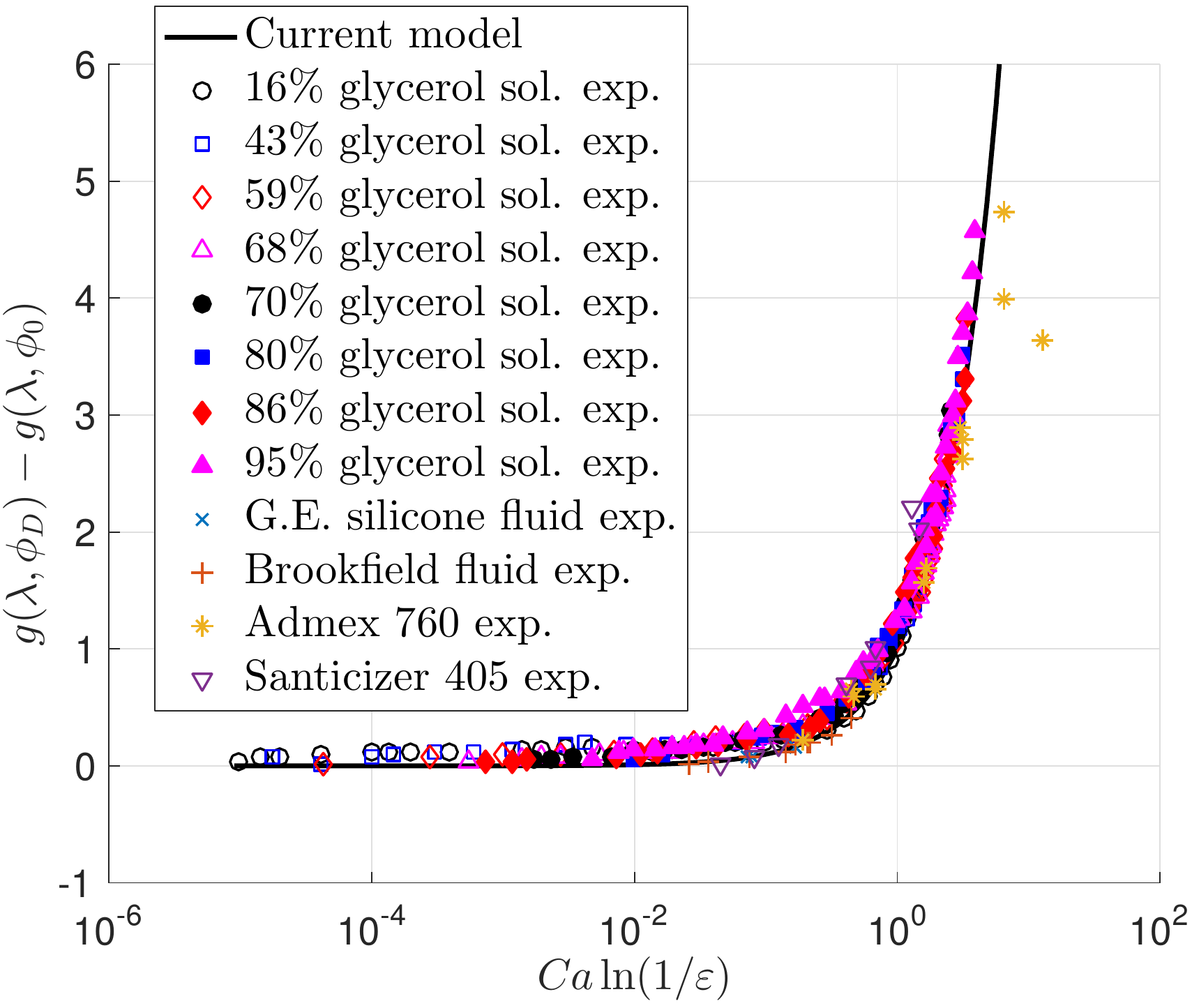}
\end{center}\end{minipage}
\caption{ Comparison of the current model to experimental data measured by Hoffman \cite{HoffmanRL:75a} and Blake \& Shikhmurzaev \cite{ShikhmurzaevYD:02a}. All glycerol solution data points were adapted from \cite{ShikhmurzaevYD:02a} while the remaining data points were adapted from \cite{HoffmanRL:75a}. To collapse the model regardless of static contact angle or viscosity ratio, the data is presented as $g(\lambda,\phi_D)-g(\lambda,\phi_0)$ vs. $\Ca \ln(1/\varepsilon)$ where $\varepsilon = 10^{-4}$. }
\label{fig:collapsed_comp}
\end{figure}

Overall, there is good agreement between the theoretical apparent dynamic contact angle and the experimental data of both Hoffman and Blake \& Shikhmurzaev. At low $\Ca$, the current model diverges slightly from the experimental data of the glycerol solution. This difference could be created by errors in contact angle measurement or by differences in the way apparent contact angle is defined. In figure~\ref{fig:collapsed_comp} we provide additional comparisons for the remaining fluids that were tested by Hoffman and Blake \& Shikhmurzaev. In this figure the magnitude of $g(\lambda,\phi_D)-g(\lambda,\phi_0)$ is plotted against $\Ca \ln(1/\varepsilon)$ so that the model can be represented by a single curve regardless of static contact angle or viscosity ratio. As before, the microscopic contact angle is predicted by equation~(\ref{eq:DCA}) and the combined model captures the dynamic contact angle behavior. There is some deviation at high $\Ca$, however this is to be expected as the current model is derived by assuming $\Ca \ll 1$. The results presented here can be extended to receding contact lines by combining our model with the model proposed by Eggers \cite{EggersJ:04c,EggersJ:05a}. One only needs to substitute negative values of $U$ into equation~(\ref{eq:DCA}) so that the advancing and receding fluids are reversed. 

In this current model, $\phi_0$ is a function of $\Ca$ and is in agreement with the experimental observations of Ram\'e et al. \cite{RameE:04a}. However, the finite force analysis does not provide a theoretical means of determining the length scale ratio, $\varepsilon$. Regardless, we found that the results are not particularly sensitive to $\varepsilon$ (in agreement with Bonn et al. \cite{BonnD:09a}) and therefore a constant value of $10^{-4}$ was used for all experimental comparisons despite the different fluid properties. This suggests that, at least for these test cases, $\varepsilon$ could be treated as a constant that is independent of the fluid properties and experimental set up.

In addition to providing a theoretical model of the dynamic contact angle, the results above will impact other aspects of MCL dynamics. For example, microscopic contact angle determines the sign and strength of vorticity near the MCL \cite{Mohseni:18f} as well as the surface tension force which is given by $\sigma \cos \phi_0$. These effects are important in microfluidics and heat/mass transfer applications where vorticity and surface tension forces influence mixing and contact line pinning. Furthermore, this result may be useful in numerical simulations where physical phenomena of interest typically span several orders of magnitude. Convergence of numerical solutions are extremely sensitive to the prescribed contact angle and grid size \cite{ZaleskiS:09a}. In general, grid convergence is achieved when the grid resolution of the simulation is the same order of magnitude as the slip length, typically $10^{-7}$m to $10^{-9}$m for water. Thus, the required number of grid points for most simulations is extremely large. In order to reduce the computational cost and improve accuracy, the Stokes flow solution could be used as a subgrid model. In such a scheme, the minimum grid size would be determined by the validity of the small $\Rey$ assumption. The implementation of a Stokes flow subgrid model is outside the scope of this publication, but will be investigated in future works.

\section{Comparison with similar singular Stokes flows}
\label{sect:force_comp}

While unfamiliar in the context of a moving contact line, a finite force corresponding to a singular stress is not unprecedented. In fact, there exist two other Stokes flows that contain stress singularities and that have known finite forces, namely the cusped interface flow and Stokeslet. The cusped interface flow, investigated by Richardson \cite{RichardsonS:68a} and Joseph et al \cite{JosephDD:90a}, is created by two submerged cylinders rotating with constant angular velocity. Under the right conditions, the fluid interface will develop a cusp singularity. The stream function near a cusped interface is reported by Richardson as
\begin{gather*}
\psi = \dfrac{\sigma}{2\pi\mu} r \ln (r) \sin(\theta).
\end{gather*}
It is easily shown that the complex formulation of this flow is given by
\begin{gather*}
G = \mu\omega + ip = -\dfrac{\sigma i}{\pi z}.
\end{gather*}
Evaluating equation~(\ref{eq:f_stress}) using Cauchy's residue theorem for a singularity that lies on the contour and at a cusp yields a force per unit length of $f = 2\sigma$, in agreement with Richardson.

A similar analysis can be performed for a two-dimensional Stokeslet, i.e. the flow that is created by an infinitely small cylinder moving through a quiescent fluid \cite{JothiramB:87a,Guazzelli:11a}. For a Stokeslet that is aligned with the $y$-axis, the stream function and complex function $G$ are given by
\begin{gather*}
\psi = \dfrac{\alpha}{4\pi} r \sin(\theta)[1-\ln (r)],\\
G = \mu\omega + ip = \dfrac{\mu \alpha i}{2\pi z},
\end{gather*}
where $\alpha$ is the strength of the Stokeslet. Evaluating equation~(\ref{eq:f_stress}) for a contour path that encloses the Stokeslet yields a force per unit length of $f = -\mu \alpha$. This result is consistent with the force reported by Avudainayagam \& Jothiram \cite{JothiramB:87a}. As discussed previously, the viscous surface force in a Stokes flow is solely determined by the pressure and vorticity. From the complex function $G$, we see that despite the different velocity fields, both the Stokeslet and the cusped fluid interface have vorticity and pressure fields that take the form of a dipole. In addition to a singular stress and finite force, we also note that the Stokeslet, cusped interface, and MCL singularity all predict infinite viscous dissipation per unit volume at $r = 0$. While this dissipation is singular, numerous applications of the Stokeslet singularity \cite{CrowdyDG:10a,Pozrikidis:90a} have demonstrated that the finite force predicted by these singular models can still be used to model physical problems.

\section{Concluding remarks}
\label{sect:conclusion}

In this publication, the force at a moving contact line was theoretically investigated using the hydrodynamic solution of the MCL. By defining a cylindrical control volume around the MCL, we were able to show using both real and complex analysis that the total viscous force exerted by the fluids on the solid is finite despite a singular stress. Unlike previous treatments of the contact line, this control volume accounts for the viscous forces that act on the fluid-fluid interface in addition to the forces that act on the fluid-solid interface, much like the Young's equation. With this finite force, we proposed a model for microscopic dynamic contact angle that is a function of the interface velocity, fluid viscositiy, and surface tension. As validation, we combined our model for microscopic contact angle with Cox's model for apparent contact angle and achieved a good match with experimental measurements. Interestingly, the results reported in this work have been alluded to in previous publications. Cox recognized the possibility of a velocity dependent microscopic contact angle and stated ``it is uncertain whether such an angle [microscopic angle] would depend on the spreading velocity'' \cite{CoxRG:86a}. In a more direct observation, Voinov \cite{VoinovOV:76a} stated ``In this case $\alpha_m$ [microscopic angle] can be a function of the velocity''. Dussan \& Davis \cite{Dussan:74a} recognized the similarities between the MCL problem and other singular models which led them state: ``There exist physical situations where the force distributed over a small area is replaced by a force acting at a point or a line. (This implies an unbounded stress tensor)". In the end, we would like to emphasize that the analysis and conclusions made in this work are for a MCL {\it model} much like the Stokeslet and cusped fluid interface. The finite force result does not imply that phenomena like slip or thermal activation do not occur, but merely that the net effect of these microscopic phenomena should yield the same total change in momentum within a microscopic control volume enclosing the contact line. 

In the field of wetting and dewetting, the concept of a finite force despite a singular stress is somewhat unusual. However, if we look to other fields, we find that there are many two-dimensional singular continuum models with similar characteristics. In electromagnetism, electric fields are singular at the locations of line charges. Through Gauss' law, we know that the strength of the point charge is finite despite the fact that the electric field is singular. In potential flow theory, line sources and line vortices are regularly used to model flows at high Reynolds numbers. Despite the fact that the velocity or shear stress approaches infinity at these singularities, they still conserve physical quantities such as mass flux and circulation. These singular models do not resolve the exact physics that occur at the singularity and are capable of correctly capturing the global features of the problem. In this sense, there may be fluid slip extremely close to the moving contact line, however we do not need to resolve it as we can already obtain the contact line force and nearby velocity field. Essentially the finite force at the moving contact line is merely a different physical application of the same mathematical concepts that are applied in other fields. At present, the MCL model retains a singular stress and infinite viscous dissipation at the corner singularity and are known limitations of this model. Motivated by the results of this work, future works will seek to extend this model to accurately capture the transfer of energy at the MCL.

\appendix
\section{General solution to the biharmonic equation}
\label{app:bih_sol}

The biharmonic stream function equation is given by
\begin{gather}
\nabla^4 \psi = 0.
\end{gather}
In polar coordinates, the solution is found using the technique of separation of variables with $\psi = R(r)\Theta(\theta)$. The solution that was initially reported by Michell \cite{MichellJH:89a}, and later extended by Filonenko-Borodich \cite{FilonenkoBM:58a}, is given by
\begin{gather}
\label{eq:psi_sol} \psi = (r/\ell)^2\ln (r/\ell) q_{2L}(\theta) + (r/\ell) \ln (r/\ell) q_{1L}(\theta) + \ln (r/\ell) q_{0L}(\theta) + \sum_{n = -\infty}^\infty (r/\ell)^n q_n(\theta).
\end{gather}
The functions $q$ are given by
\begin{align*}
q_{2L} &= P_{2L}[A_{2L}+B_{2L}\theta],\\
q_{1L} &= P_{1L}[A_{1L} \cos(\theta)+ B_{1L} \sin(\theta) +C_{1L} \theta \cos(\theta) +D_{1L} \theta \sin(\theta)],\\
q_{0L} &= P_{0L}[A_{0L} + B_{0L}\theta],\\
q_0 &= P_0[A_{0} + B_{0} \theta + C_{0} \cos(2\theta) + D_{0} \sin(2\theta)],\\
q_1 &= P_1[A_{1} \cos(\theta) + B_{1} \sin(\theta) + C_{1} \theta \cos(\theta) + D_{1} \theta \sin(\theta)],\\
q_2 &= P_2[A_{2} + B_{2} \theta + C_{2} \cos(2\theta) + D_{2} \sin(2\theta)],\\
q_n & = P_n[A_{n} \cos((n-2)\theta) +B_{n} \cos(n\theta) +C_{n}\sin((n-2)\theta) + D_{n} \sin(n\theta)] \quad \text{for} \quad n \geq 3,\\
q_n &= P_n[A_{n} \cos ((n+2)\theta) +B_{n} \cos(n\theta) + C_{n} \sin((n+2)\theta) + D_{n} \sin(n\theta)] \quad \text{for} \quad n\leq -1,
\end{align*}
where $P$ is a dimensional coefficient and $A$, $B$, $C$, and $D$ are coefficients determined by the boundary conditions. The velocity, vorticity, and pressure can all be determined using the equations given by
\begin{gather}
\bm{u} = \bnabla \times \psi \bm{\hat{e}_z},\\
\omega = -\bnabla^2 \psi,\\
\bnabla p = \mu\nabla^2 \bm{u}.
\end{gather}

In the past, stream functions of various order $r$ have been investigated individually and correlated to unique classes of flows. The paint scraper problem investigated by Taylor \cite{Taylor:62a} is described terms of order $(r/\ell)^{-1}$. Similarly, the MCL problem investigated by Huh \& Scriven \cite{HuhC:71a} is described by terms of order $(r/\ell)^{-1}$. Moffatt eddies and the hinged plate flow correspond to terms with $n \geq 1$ \cite{MoffattHK:64a}. The cusped fluid interface and Stokeslet solution described by Richardson \cite{RichardsonS:68a} and Joseph et al \cite{JosephDD:90a} is described by terms of order $(r/\ell) \ln (r/\ell)$. Solutions that combine multiple orders of $n$ have been shown to represent more complex flows such as evaporation near the contact line \cite{SnoeijerJH:12a,SnoeijerJH:13a}.

\section{Zeroth order solution to the moving contact line flow}
\label{app:coeffs}
The zeroth order solution to the moving contact line problem described in \S\ref{sect:geo_sol} was first reported by Moffatt \cite{MoffattHK:64a} and Huh \& Scriven \cite{HuhC:71a}.  In a planar wedge geometry, where fluid A has a contact angle of $\phi_0$, the flow is governed by the biharmonic stream function equations given by
\begin{gather*}
\nabla^4 \psi_{A0} = 0,\\
\nabla^4 \psi_{B0} = 0.
\end{gather*}
The boundary conditions of the zeroth order solution include no-slip at the fluid-solid interface, zero mass flux through all interfaces, and continuity of tangential velocity and shear stress across the fluid-fluid interface. The resulting boundary conditions on $\psi_{A0}$ and $\psi_{B0}$ are therefore given by
\begin{equation*}
\begin{aligned}[c]
\psi_{A0}(\theta = 0) &= 0, \\
\psi_{B0}(\theta = \phi_0) &= 0, \\
\left[ \dfrac{1}{r} \dfrac{\partial \psi_{A0}}{\partial \theta} \right]_{\theta = 0} &= U,\\
\left[  \dfrac{\partial \psi_{A0}}{\partial \theta} \right]_{\theta = \phi_0} &= \left[  \dfrac{\partial \psi_{B0}}{\partial \theta} \right]_{\theta = \phi_0},
\end{aligned}
\hspace{3 cm}
\begin{aligned}[c]
\psi_{A0}(\theta = \phi_0) &= 0,\\
\psi_{B0}(\theta = \pi) &= 0,\\
\left[ \dfrac{1}{r} \dfrac{\partial \psi_{B0}}{\partial \theta} \right]_{\theta = \pi} &= -U,\\
 \left[ \dfrac{\partial^2 \psi_{A0}}{\partial \theta^2} \right]_{\theta = \phi_0} &=  \left[ \lambda \dfrac{\partial^2 \psi_{B0}}{\partial \theta^2} \right]_{\theta = \phi_0},
\end{aligned}
\end{equation*}
where $\lambda = \mu_B/\mu_A$ denotes the viscosity ratio. Solving the system of equations created by the boundary conditions above yields the stream function, velocity, pressure, and vorticity is given by
\begin{gather*}
\psi_0 =rU[A\cos(\theta) + B \sin(\theta) + C\theta\cos(\theta)+D\theta \sin(\theta)],\\
u_{0r} = \dfrac{1}{r} \dfrac{\partial \psi}{\partial \theta} =U\left[-A\sin(\theta) + B\cos(\theta) + C[\cos(\theta)-\theta\sin(\theta)] + D[\sin(\theta) + \theta\cos(\theta)] \dfrac{}{} \right],\\
u_{0\theta} = -\dfrac{\partial \psi}{\partial r} = -U[A\cos(\theta)+B\sin(\theta)+C\theta\cos(\theta)+D\theta\sin(\theta)],\\
p_0 = \dfrac{2\mu U}{r} [C\cos(\theta)+D\sin(\theta)],\\
\omega_0 = \dfrac{2U}{r} [C\sin(\theta) - D\cos(\theta)].
\end{gather*}
For each fluid $A$, $B$, $C$, and $D$ are constant coefficients given by
\begin{align*}
A_A &= 0,\\
B_A &= \dfrac{(8\phi_0(\lambda - 1))\sin(\phi_0)^2 + 4\pi\lambda\phi_0\sin(2\phi_0) + 8\phi_0(\pi^2 - \lambda\phi_0^2 - 2\pi\phi_0 + \phi_0^2 + \pi\lambda\phi_0)}{\Delta},\\
C_A &= [-8\pi\lambda\sin(\phi_0)^2 + (8\pi\phi_0 - 2\lambda + 4\lambda\phi_0^2 - 4\pi^2 - 4\phi_0^2 - 4\pi\lambda\phi_0 + 2)\sin(2\phi_0) \\
& \quad+ \lambda\sin(4\phi_0) - \sin(4\phi_0)]/\Delta,\\
D_A &= \dfrac{(8 - 8\lambda)\sin(\phi_0)^4 + (16\pi\phi_0 + 8\lambda\phi_0^2 - 8\pi^2 - 8\phi_0^2 - 8\pi\lambda\phi_0)\sin(\phi_0)^2}{\Delta}, 
\end{align*}
\begin{align*}
A_B &= \dfrac{8\pi^2\sin(\phi_0)^2 + (-\pi(4\pi\phi_0 - 2\lambda + 4\lambda\phi_0^2 - 4\phi_0^2 + 2))\sin(2\phi_0) + \pi(\sin(4\phi_0) - \lambda\sin(4\phi_0))}{\Delta},\\
B_B &= [(8\lambda\phi_0 - 8\phi_0 - 8\phi_0\pi^2 + 8\pi\phi_0^2 - 8\pi\lambda\phi_0^2)\sin(\phi_0)^2 + (2\pi - 2\pi\lambda)\sin(2\phi_0)^2 \\
& \quad + (4\pi\phi_0 - 4\pi^2)\sin(2\phi_0) + 8\phi_0\pi^2 - 16\pi\phi_0^2 - 8\lambda\phi_0^3 + 8\phi_0^3 + 8\pi\lambda\phi_0^2]/\Delta,\\
C_B &= \dfrac{-8\pi\sin(\phi_0)^2 + (4\pi\phi_0 - 2\lambda + 4\lambda\phi_0^2 - 4\phi_0^2 + 2)\sin(2\phi_0) + \lambda\sin(4\phi_0) - \sin(4\phi_0)}{\Delta},\\
D_B &= \dfrac{(8 - 8\lambda)\sin(\phi_0)^4 + (8\pi\phi_0 + 8\lambda\phi_0^2 - 8\phi_0^2)\sin(\phi_0)^2}{\Delta},\\
\Delta &= (\phi_0(8\lambda - 8) - 8\pi\lambda)\sin(\phi_0)^2 + ((4\lambda - 4)\phi_0^2 + 8\pi\phi_0 - 2\lambda - 4\pi^2 + 2)\sin(2\phi_0)\\
& \quad + (\lambda - 1)\sin(4\phi_0) + (8 - 8\lambda)\phi_0^3 + (8\lambda\pi - 16\pi)\phi_0^2 + 8\pi^2\phi_0.
\end{align*}
Note that the coefficients are a function of the contact angle, $\phi_0$, and viscosity ratio, $\lambda$, only. 


\section{Complex contour integrals with singularities on the contour}
\label{app:SP_theorem}

The contour integrals considered in this paper typically contain singularities that are simultaneously on the contour and at a corner. In this section, we demonstrate how Cauchy's residue theorem is affected by singularities on the boundary based on \S1.5 of \cite{EstradaR:12a}. 

\begin{figure}
\centering
\begin{minipage}{0.24\linewidth}\begin{center}
 \includegraphics[width=0.69\linewidth]{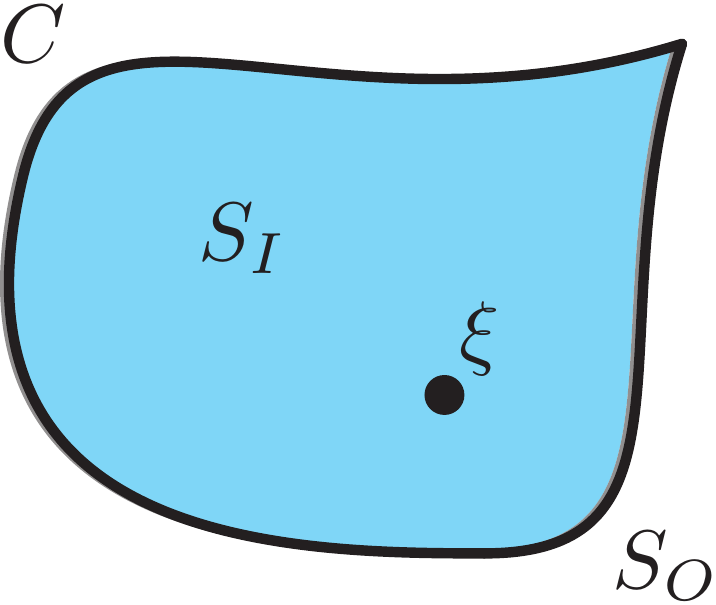}
\end{center}\end{minipage}
\begin{minipage}{0.24\linewidth}\begin{center}
 \includegraphics[width=0.69\linewidth]{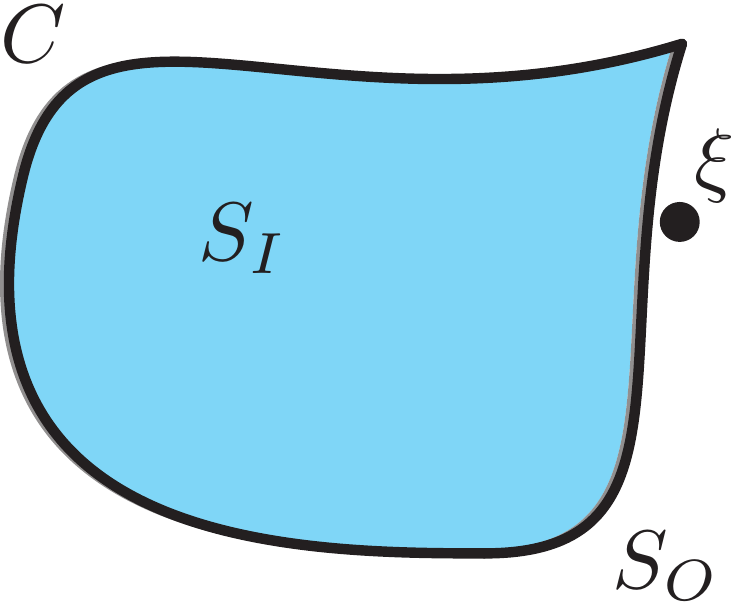}
\end{center}\end{minipage}
\begin{minipage}{0.24\linewidth}\begin{center}
 \includegraphics[width=0.69\linewidth]{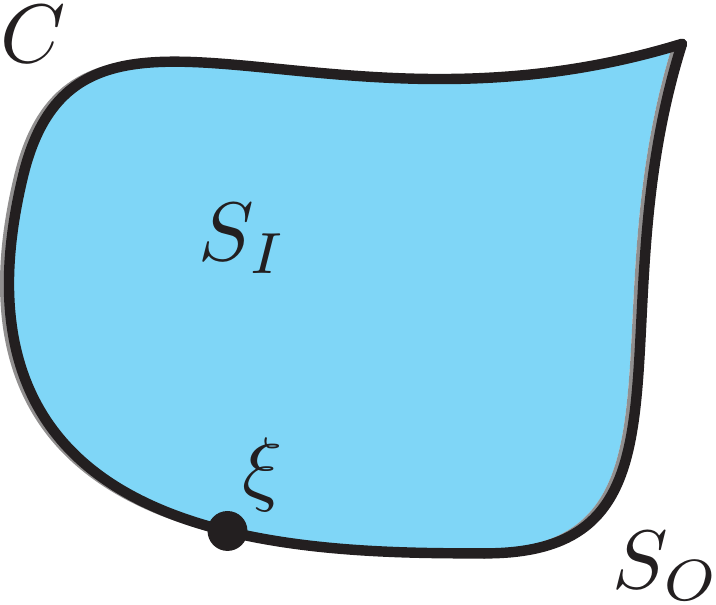}
\end{center}\end{minipage}
\begin{minipage}{0.24\linewidth}\begin{center}
 \includegraphics[width=0.69\linewidth]{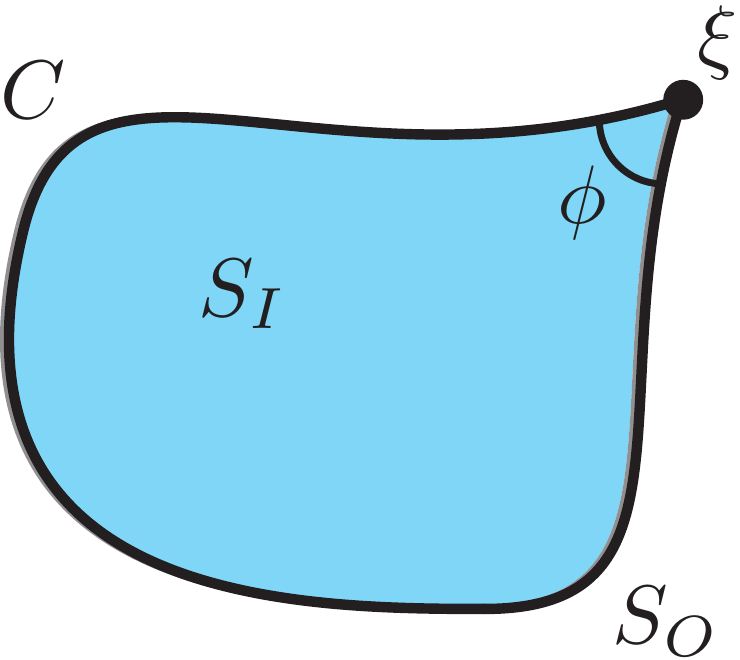}
\end{center}\end{minipage}
\begin{minipage}{0.23\linewidth}\begin{center}
(a) $\xi \in S_I$
\end{center}\end{minipage}
\begin{minipage}{0.24\linewidth}\begin{center}
(b) $\xi \in S_O$
\end{center}\end{minipage}
\begin{minipage}{0.24\linewidth}\begin{center}
(c) $\xi \in C$
\end{center}\end{minipage}
\begin{minipage}{0.26\linewidth}\begin{center}
(d) $\xi \in C$ and at a corner
\end{center}\end{minipage}
\caption{(a) $\xi$ lies inside the contour in the region $S_I$ (b) $\xi$ lies outside the contour in the region $S_O$ (c) $\xi$ lies on the path $C$ along a smooth segment (d) $\xi$ lies on the path $C$ at a corner point.}
\label{fig:sokhotski}
\end{figure}

Consider the analytic function $W = 1/(z-\xi)$ and the closed contour $C$ that defines an interior region, $S_I$, and an exterior region, $S_O$. The solution to the complex contour integral of $W$ along the path $C$ is given by
	\begin{gather*}
		\oint_C W dz=\begin{cases}
		2\pi i, & \xi \in S_I,\\
    		0, & \xi \in S_O,\\
		\pi i, & \xi \in C \text{ along a smooth segment of C},\\
		\phi i, & \xi \in C \text{ and at a corner},
  		\end{cases}
	\end{gather*} 
where the four cases correspond to the location of $\xi$, as shown in figure~\ref{fig:sokhotski}. In the first two cases, the singularity lies inside the contour ($\xi \in S_I$) or outside the contour ($\xi \in S_O$) and Cauchy's residue theorem yields a solution of $2\pi i\Res(W)$ and 0, respectively. If $\xi$ lies on a smooth segment of $C$, then the Sokhotski-Plemelj theorem states that the integral will equal $\pi i\Res(W)$, which in this case is $\pi i$. If $\xi$ lies on $C$ and at a corner, then the Sokhotski-Plemelj theorem states that the integral will equal $\phi i\Res(W) = \phi i$, where $\phi$ is the angle of the corner. 

\section*{References}

\bibliography{/Users/Peter/GitHub/REF/RefA2,/Users/Peter/GitHub/REF/ref_zhang}

\begin{thebibliography}{10}
\expandafter\ifx\csname url\endcsname\relax
  \def\url#1{\texttt{#1}}\fi
\expandafter\ifx\csname urlprefix\endcsname\relax\def\urlprefix{URL }\fi
\expandafter\ifx\csname href\endcsname\relax
  \def\href#1#2{#2} \def\path#1{#1}\fi

\bibitem{YarinAL:06a}
A.~L. Yarin,
  \href{http://dx.doi.org/10.1146/annurev.fluid.38.050304.092144}{Drop impact
  dynamics: {Splashing}, spreading, receding, bouncing{...}}, Annual Review of
  Fluid Mechanics 38 (2006) 159--192.
\newblock \href {http://dx.doi.org/10.1146/annurev.fluid.38.050304.092144}
  {\path{doi:10.1146/annurev.fluid.38.050304.092144}}.
\newline\urlprefix\url{http://dx.doi.org/10.1146/annurev.fluid.38.050304.092144}

\bibitem{DhirVK:98a}
V.~K. Dhir, \href{http://dx.doi.org/10.1146/annurev.fluid.30.1.365}{Boiling
  heat transfer}, Annual Review of Fluid Mechanics 30~(1) (1998) 365--401.
\newblock \href {http://dx.doi.org/10.1146/annurev.fluid.30.1.365}
  {\path{doi:10.1146/annurev.fluid.30.1.365}}.
\newline\urlprefix\url{http://dx.doi.org/10.1146/annurev.fluid.30.1.365}

\bibitem{WeinsteinS:04a}
S.~J. Weinstein, K.~J. Ruschak, Coating flows, Annu. Rev. Fluid Mech. 36 (2004)
  29--53.

\bibitem{DerbyB:10a}
B.~Derby, Inkjet printing of functional and structural materials: fluid
  property requirements, feature stability, and resolution, Annual Review of
  Materials Research 40 (2010) 395--414.

\bibitem{SnoeijerJH:13b}
J.~H. Snoeijer, B.~Andreotti,
  \href{http://dx.doi.org/10.1146/annurev-fluid-011212-140734}{Moving contact
  lines: {S}cales, regimes, and dynamical transitions}, Annual Review of Fluid
  Mechanics 45 (2013) 269--292.
\newblock \href {http://dx.doi.org/10.1146/annurev-fluid-011212-140734}
  {\path{doi:10.1146/annurev-fluid-011212-140734}}.
\newline\urlprefix\url{http://dx.doi.org/10.1146/annurev-fluid-011212-140734}

\bibitem{Dussan:76a}
E.~B. Dussan, \href{http://dx.doi.org/10.1017/s0022112076002838}{The moving
  contact line: the slip boundary condition}, Journal of Fluid Mechanics 77~(4)
  (1976) 665--684.
\newblock \href {http://dx.doi.org/10.1017/s0022112076002838}
  {\path{doi:10.1017/s0022112076002838}}.
\newline\urlprefix\url{http://dx.doi.org/10.1017/s0022112076002838}

\bibitem{EralH:13a}
H.~Eral, J.~Oh, et~al., Contact angle hysteresis: a review of fundamentals and
  applications, Colloid and polymer science 291~(2) (2013) 247--260.

\bibitem{QuereD:08a}
D.~Qu{\'e}r{\'e}, Wetting and roughness, Annu. Rev. Mater. Res. 38 (2008)
  71--99.

\bibitem{deGennesPG:85a}
P.-G. {de Gennes}, \href{http://dx.doi.org/10.1103/RevModPhys.57.827}{Wetting:
  {Statistics} and dynamics}, Reviews of Modern Physics 57 (1985) 827--863.
\newblock \href {http://dx.doi.org/10.1103/RevModPhys.57.827}
  {\path{doi:10.1103/RevModPhys.57.827}}.
\newline\urlprefix\url{http://dx.doi.org/10.1103/RevModPhys.57.827}

\bibitem{LandauL:88a}
L.~Landau, B.~Levich, Dragging of a liquid by a moving plate, in: Dynamics of
  Curved Fronts, Elsevier, 1988, pp. 141--153.

\bibitem{KumarS:14a}
E.~Vandre, M.~Carvalho, S.~Kumar, Characteristics of air entrainment during
  dynamic wetting failure along a planar substrate, Journal of Fluid Mechanics
  747 (2014) 119--140.

\bibitem{MugeleF:05a}
F.~Mugele, J.~C. Baret,
  \href{http://dx.doi.org/10.1088/0953-8984/17/28/R01}{Electrowetting: {From}
  basics to applications}, Journal of Physics: Condensed Matter 17~(28) (2005)
  705--774.
\newblock \href {http://dx.doi.org/10.1088/0953-8984/17/28/R01}
  {\path{doi:10.1088/0953-8984/17/28/R01}}.
\newline\urlprefix\url{http://dx.doi.org/10.1088/0953-8984/17/28/R01}

\bibitem{SnoeijerJH:12a}
H.~Gelderblom, O.~Bloemen, J.~Snoeijer, Stokes flow near the contact line of an
  evaporating drop, Journal of fluid mechanics 709 (2012) 69--84.

\bibitem{SnoeijerJH:13a}
H.~Gelderblom, H.~Stone, J.~Snoeijer, Stokes flow in a drop evaporating from a
  liquid subphase, Physics of Fluids (1994-present) 25~(10) (2013) 102102.

\bibitem{SpeltP:14a}
Y.~Sui, H.~Ding, P.~D.~M. Spelt,
  \href{https://doi.org/10.1146/annurev-fluid-010313-141338}{Numerical
  simulations of flows with moving contact lines}, Annual Review of Fluid
  Mechanics 46 (2014) 97--119.
\newblock \href {http://dx.doi.org/10.1146/annurev-fluid-010313-141338}
  {\path{doi:10.1146/annurev-fluid-010313-141338}}.
\newline\urlprefix\url{https://doi.org/10.1146/annurev-fluid-010313-141338}

\bibitem{BlakeTD:06a}
T.~D. Blake, \href{https://doi.org/10.1016/j.jcis.2006.03.051}{The physics of
  moving wetting lines}, Journal of Colloid and Interface Science 299~(1)
  (2006) 1--13.
\newblock \href {http://dx.doi.org/10.1016/j.jcis.2006.03.051}
  {\path{doi:10.1016/j.jcis.2006.03.051}}.
\newline\urlprefix\url{https://doi.org/10.1016/j.jcis.2006.03.051}

\bibitem{BonnD:09a}
D.~Bonn, J.~Eggers, J.~Indekeu, J.~Meunier, E.~Rolley,
  \href{http://dx.doi.org/10.1103/RevModPhys.81.739}{Wetting and spreading},
  Reviews of modern physics 81 (2009) 739--805.
\newblock \href {http://dx.doi.org/10.1103/RevModPhys.81.739}
  {\path{doi:10.1103/RevModPhys.81.739}}.
\newline\urlprefix\url{http://dx.doi.org/10.1103/RevModPhys.81.739}

\bibitem{MoffattHK:64a}
H.~K. Moffatt, Viscous and resistive eddies near a sharp corner, Journal of
  Fluid Mechanics 18~(01) (1964) 1--18.

\bibitem{HuhC:71a}
C.~Huh, L.~E. Scriven,
  \href{http://dx.doi.org/10.1016/0021-9797(71)90188-3}{Hydrodynamic model of
  steady movement of a solid/liquid/fluid contact line}, Journal of Colloid and
  Interface Science 35~(1) (1971) 85--101.
\newblock \href {http://dx.doi.org/10.1016/0021-9797(71)90188-3}
  {\path{doi:10.1016/0021-9797(71)90188-3}}.
\newline\urlprefix\url{http://dx.doi.org/10.1016/0021-9797(71)90188-3}

\bibitem{HockingLM:77a}
L.~M. Hocking, A moving fluid interface. {Part} 2. {The} removal of the force
  singularity by a slip flow, Journal of Fluid Mechanics 79 (1977) 209--229.

\bibitem{CoxRG:86a}
R.~G. Cox, \href{http://dx.doi.org/10.1017/S0022112086000332}{The dynamics of
  the spreading of liquids on a solid surface. {Part} 1. {Viscous} flow},
  Journal of Fluid Mechanics 168 (1986) 169--194.
\newblock \href {http://dx.doi.org/10.1017/S0022112086000332}
  {\path{doi:10.1017/S0022112086000332}}.
\newline\urlprefix\url{http://dx.doi.org/10.1017/S0022112086000332}

\bibitem{ShikhmurzaevYD:97a}
Y.~D. Shikhmurzaev, Moving contact lines in liquid/liquid/solid systems,
  Journal of Fluid Mechanics 334 (1997) 211249.

\bibitem{EggersJ:04a}
J.~Eggers, H.~A. Stone,
  \href{https://doi.org/10.1017/S0022112004008663}{Characteristic lengths at
  moving contact lines for a perfectly wetting fluid: the influence of speed on
  the dynamic contact angle}, Journal of Fluid Mechanics 505 (2004) 309--321.
\newblock \href {http://dx.doi.org/10.1017/S0022112004008663}
  {\path{doi:10.1017/S0022112004008663}}.
\newline\urlprefix\url{https://doi.org/10.1017/S0022112004008663}

\bibitem{BlakeTD:69a}
T.~D. Blake, J.~M. Haynes,
  \href{https://doi.org/10.1016/0021-9797(69)90411-1}{Kinetics of
  {liquid/liquid} displacement}, Journal of Colloid and Interface Science
  30~(3) (1969) 421--423.
\newblock \href {http://dx.doi.org/10.1016/0021-9797(69)90411-1}
  {\path{doi:10.1016/0021-9797(69)90411-1}}.
\newline\urlprefix\url{https://doi.org/10.1016/0021-9797(69)90411-1}

\bibitem{ShikhmurzaevYD:07a}
Y.~D. Shikhmurzaev, Capillary Flows with Forming Interfaces, CRC Press, 2007.

\bibitem{VoinovOV:76a}
O.~V. Voinov, \href{http://dx.doi.org/10.1007/bf01012963}{Hydrodynamics of
  wetting}, Fluid Dynamics 11~(5) (1976) 714--721.
\newblock \href {http://dx.doi.org/10.1007/bf01012963}
  {\path{doi:10.1007/bf01012963}}.
\newline\urlprefix\url{http://dx.doi.org/10.1007/bf01012963}

\bibitem{PetrovP:92a}
P.~Petrov, I.~Petrov, A combined molecular-hydrodynamic approach to wetting
  kinetics, Langmuir 8~(7) (1992) 1762--1767.

\bibitem{PismenL:02a}
L.~Pismen, Mesoscopic hydrodynamics of contact line motion, Colloids and
  Surfaces A: Physicochemical and Engineering Aspects 206~(1-3) (2002) 11--30.

\bibitem{SeppecherP:96a}
P.~Seppecher, \href{http://dx.doi.org/10.1016/0020-7225(95)00141-7}{Moving
  contact lines in the {Cahn-Hilliard} theory}, Journal of Computational
  Physics 34~(9) (1996) 977--992.
\newblock \href {http://dx.doi.org/10.1016/0020-7225(95)00141-7}
  {\path{doi:10.1016/0020-7225(95)00141-7}}.
\newline\urlprefix\url{http://dx.doi.org/10.1016/0020-7225(95)00141-7}

\bibitem{ShikhmurzaevYD:06a}
Y.~D. Shikhmurzaev,
  \href{http://dx.doi.org/10.1016/j.physd.2006.03.003}{Singularities at the
  moving contact line. {Mathematical}, physical and computational aspects},
  Physica D: Nonlinear Phenomena 217~(2) (2006) 121--133.
\newblock \href {http://dx.doi.org/10.1016/j.physd.2006.03.003}
  {\path{doi:10.1016/j.physd.2006.03.003}}.
\newline\urlprefix\url{http://dx.doi.org/10.1016/j.physd.2006.03.003}

\bibitem{NavierCL:23a}
C.~L. M.~H. Navier, Memoire sur les lois du mouvement des fluides, in: Memoires
  de l Academie Royale des Sciences de l Instituede France, Vol.~6, Royale des
  Sciences l Instituede France, 1823, pp. 389--440.

\bibitem{Troian:97a}
P.~A. Thompson, S.~M. Troian, \href{http://dx.doi.org/10.1038/38686}{A general
  boundary condition for liquid flow at solid surfaces}, Nature 389~(6649)
  (1997) 360--362.
\newblock \href {http://dx.doi.org/10.1038/38686} {\path{doi:10.1038/38686}}.
\newline\urlprefix\url{http://dx.doi.org/10.1038/38686}

\bibitem{Mohseni:16b}
J.~J. Thalakkottor, K.~Mohseni,
  \href{http://dx.doi.org/10.1103/PhysRevE.94.023113}{Universal slip boundary
  condition for fluid flows}, Physical Review {E} 94 (2016) 023113.
\newblock \href {http://dx.doi.org/10.1103/PhysRevE.94.023113}
  {\path{doi:10.1103/PhysRevE.94.023113}}.
\newline\urlprefix\url{http://dx.doi.org/10.1103/PhysRevE.94.023113}

\bibitem{HockingLM:82a}
L.~M. Hocking, A.~D. Rivers,
  \href{http://dx.doi.org/10.1017/S0022112082001979}{The spreading of a drop by
  capillary action}, Journal of Fluid Mechanics 121 (1982) 425--442.
\newblock \href {http://dx.doi.org/10.1017/S0022112082001979}
  {\path{doi:10.1017/S0022112082001979}}.
\newline\urlprefix\url{http://dx.doi.org/10.1017/S0022112082001979}

\bibitem{RameE:96a}
E.~Ram{\'e}, S.~Garoff, Microscopic and macroscopic dynamic interface shapes
  and the interpretation of dynamic contact angles, Journal of colloid and
  interface science 177~(1) (1996) 234--244.

\bibitem{SuiY:13a}
Y.~Sui, P.~D. Spelt, Validation and modification of asymptotic analysis of slow
  and rapid droplet spreading by numerical simulation, Journal of Fluid
  Mechanics 715 (2013) 283--313.

\bibitem{RameE:04a}
E.~Ram{\'e}, S.~Garoff, K.~Willson,
  \href{http://dx.doi.org/10.1103/PhysRevE.70.031608}{Characterizing the
  microscopic physics near moving contact lines using dynamic contact angle
  data}, Physical Review E 70 (2004) 031608.
\newblock \href {http://dx.doi.org/10.1103/PhysRevE.70.031608}
  {\path{doi:10.1103/PhysRevE.70.031608}}.
\newline\urlprefix\url{http://dx.doi.org/10.1103/PhysRevE.70.031608}

\bibitem{ShenC:98a}
C.~Shen, D.~W. Ruth, Experimental and numerical investigations of the interface
  profile close to a moving contact line, Physics of Fluids 10~(4) (1998)
  789--799.

\bibitem{ZhouM:92a}
P.~Sheng, M.~Zhou, Immiscible-fluid displacement: Contact-line dynamics and the
  velocity-dependent capillary pressure, Physical review A 45~(8) (1992) 5694.

\bibitem{HoffmanRL:75a}
R.~L. Hoffman, A study of the advancing interface. i. interface shape in
  liquid-gas systems, Journal of colloid and interface science 50~(2) (1975)
  228--241.

\bibitem{ShikhmurzaevYD:02a}
T.~Blake, Y.~D. Shikhmurzaev, Dynamic wetting by liquids of different
  viscosity, Journal of Colloid and Interface Science 253~(1) (2002) 196--202.

\bibitem{SevenoD:09a}
D.~Seveno, A.~Vaillant, R.~Rioboo, H.~Adao, J.~Conti, J.~D. Coninck, Dynamics
  of wetting revisited, Langmuir 25~(22) (2009) 13034--13044.

\bibitem{BlakeT:11a}
D.~Duvivier, D.~Seveno, R.~Rioboo, T.~Blake, J.~D. Coninck, Experimental
  evidence of the role of viscosity in the molecular kinetic theory of dynamic
  wetting, Langmuir 27~(21) (2011) 13015--13021.

\bibitem{ItoT:15a}
K.~Katoh, T.~Wakimoto, Y.~Yamamoto, T.~Ito, Dynamic wetting behavior of a
  triple-phase contact line in several experimental systems, Experimental
  Thermal and Fluid Science 60 (2015) 354--360.

\bibitem{Dussan:79a}
E.~B. Dussan, \href{http://dx.doi.org/10.1146/annurev.fl.11.010179.002103}{On
  the spreading of liquids on solid surfaces: {Static} and dynamic contact
  lines}, Annual Review of Fluid Mechanics 11 (1979) 371--400.
\newblock \href {http://dx.doi.org/10.1146/annurev.fl.11.010179.002103}
  {\path{doi:10.1146/annurev.fl.11.010179.002103}}.
\newline\urlprefix\url{http://dx.doi.org/10.1146/annurev.fl.11.010179.002103}

\bibitem{GriffithsDJ:72a}
D.~J. Griffiths, Introduction to Electrodynamics, Prentice Hall, Englewood
  Cliffs, NJ, USA, 1972.

\bibitem{Batchelor:67a}
G.~K. Batchelor, An Introduction to Fluid Dynamics, Cambridge University Press,
  Cambridge, UK, 1967.

\bibitem{CrowdyDG:10a}
D.~Crowdy, Y.~Or, Two-dimensional point singularity model of a
  {low-Reynolds-number} swimmer near a wall, Physical Review E 81~(3) (2010)
  036313.

\bibitem{SnoeijerJ:06a}
J.~H. Snoeijer, Free-surface flows with large slopes: Beyond lubrication
  theory, Physics of Fluids 18~(2) (2006) 021701.

\bibitem{SibleyD:15a}
D.~N. Sibley, A.~Nold, S.~Kalliadasis, The asymptotics of the moving contact
  line: cracking an old nut, Journal of Fluid Mechanics 764 (2015) 445--462.

\bibitem{SlatteryJC:07a}
J.~C. Slattery, L.~Sagis, E.~S. Oh, Interfacial Transport Phenomena, 2nd
  Edition, Springer, New York, New York, NY, USA, 2007.

\bibitem{mohseni:17r}
J.~J. Thalakkottor, K.~Mohseni, The role of surface tension gradient in
  determining microscopic dynamic contact angle, Journal of Fluid
  MechanicsSubmitted.

\bibitem{WuJZ:06a}
J.-Z. Wu, H.-Y. Ma, M.-D. Zhou, Vorticity and vortex dynamics, Springer, 2006.

\bibitem{Dussan:74a}
E.~B. Dussan, S.~H. Davis, \href{https://doi.org/10.1017/S0022112074001261}{On
  the motion of a fluid-fluid interface along a solid surface}, Journal of
  Fluid Mechanics 65~(1) (1974) 71--95.
\newblock \href {http://dx.doi.org/10.1017/S0022112074001261}
  {\path{doi:10.1017/S0022112074001261}}.
\newline\urlprefix\url{https://doi.org/10.1017/S0022112074001261}

\bibitem{Guazzelli:11a}
E.~Guazzelli, J.~F. Morris, A physical introduction to suspension dynamics,
  Cambridge University Press, 2011.

\bibitem{LangloisWE:64a}
W.~Langlois, M.~Deville, Slow viscous flow, Springer, 1964.

\bibitem{MitrinovicD:84a}
D.~Mitrinovic, J.~Keckic, The Cauchy method of residues: theory and
  applications, Vol. 259, Springer Science \& Business Media, 1984.

\bibitem{EstradaR:12a}
R.~Estrada, R.~Kanwal, Singular integral equations, Springer Science \&
  Business Media, 2012.

\bibitem{CrowdyD:17a}
D.~G. Crowdy, S.~J. Brzezicki, Analytical solutions for two-dimensional stokes
  flow singularities in a no-slip wedge of arbitrary angle, Proc. R. Soc. A
  473~(2202) (2017) 20170134.

\bibitem{LealLG:07a}
L.~G. Leal, Advanced transport phenomena: fluid mechanics and convective
  transport processes, Cambridge University Press, New York, NY, USA, 2007.

\bibitem{SnoeijerJH:16a}
B.~Andreotti, J.~Snoeijer, Soft wetting and the shuttleworth effect, at the
  crossroads between thermodynamics and mechanics, Europhysics Letters 113~(6)
  (2016) 66001.

\bibitem{JonesMA:03a}
M.~A. Jones, \href{http://dx.doi.org/10.1017/S0022112003006645}{The separated
  flow of an inviscid fluid around a moving flat plate}, Journal of Fluid
  Mechanics 496 (2003) 405--441.
\newblock \href {http://dx.doi.org/10.1017/S0022112003006645}
  {\path{doi:10.1017/S0022112003006645}}.
\newline\urlprefix\url{http://dx.doi.org/10.1017/S0022112003006645}

\bibitem{LesterG:61a}
G.~Lester, Contact angles of liquids at deformable solid surfaces, Journal of
  Colloid Science 16~(4) (1961) 315--326.

\bibitem{EW:07a}
W.~Ren, W.~E, \href{http://dx.doi.org/10.1063/1.2646754}{Boundary conditions
  for the moving contact line problem}, Physics of Fluids 19~(2) (2007)
  022101(1--15).
\newblock \href {http://dx.doi.org/10.1063/1.2646754}
  {\path{doi:10.1063/1.2646754}}.
\newline\urlprefix\url{http://dx.doi.org/10.1063/1.2646754}

\bibitem{BlakeTD:11a}
T.~D. Blake, J.~D. Coninck, Dynamics of wetting and kramers’ theory, The
  European Physical Journal Special Topics 197~(1) (2011) 249.

\bibitem{EggersJ:04c}
J.~Eggers, Hydrodynamic theory of forced dewetting, Physical review letters
  93~(9) (2004) 094502.

\bibitem{EggersJ:05a}
J.~Eggers, Existence of receding and advancing contact lines, Physics of Fluids
  17~(8) (2005) 082106.

\bibitem{Mohseni:18f}
P.~Zhang, K.~Mohseni, Dipole model of vorticity at the moving contact line,
  International Journal of Multiphase Flow 103 (2018) 169--172.
\newblock \href
  {http://dx.doi.org/https://doi.org/10.1016/j.ijmultiphaseflow.2018.02.008}
  {\path{doi:https://doi.org/10.1016/j.ijmultiphaseflow.2018.02.008}}.

\bibitem{ZaleskiS:09a}
S.~Afkhami, S.~Zaleski, M.~Bussmann, A mesh-dependent model for applying
  dynamic contact angles to {VOF} simulations, Journal of Computational Physics
  228~(15) (2009) 5370--5389.

\bibitem{RichardsonS:68a}
S.~Richardson, Two-dimensional bubbles in slow viscous flows, Journal of Fluid
  Mechanics 33~(03) (1968) 475--493.

\bibitem{JosephDD:90a}
D.~D. Joseph, J.~Nelson, M.~Renardy, Y.~Renardy, Two-dimensional cusped
  interfaces, Journal of fluid mechanics 223 (1990) 383--409.

\bibitem{JothiramB:87a}
A.~Avudainayagam, B.~Jothiram, No-slip images of certain line singularities in
  a circular cylinder, International journal of engineering science 25~(9)
  (1987) 1193--1205.

\bibitem{Pozrikidis:90a}
C.~Pozrikidis, The axisymmetric deformation of a red blood cell in uniaxial
  straining {Stokes} flow, Journal of Fluid Mechanics 216 (1990) 231--254.

\bibitem{MichellJH:89a}
J.~Michell, On the direct determination of stress in an elastic solid, with
  application to the theory of plates, Proceedings of the London Mathematical
  Society 1~(1) (1899) 100--124.

\bibitem{FilonenkoBM:58a}
M.~Filonenko-Borodich, Theory of Elasticity, University Press of the Pacific,
  Moscow, 1958.

\bibitem{Taylor:62a}
G.~Taylor, On scraping viscous fluid from a plane surface, Miszellangen der
  Angewandten Mechanik (Festschrift Walter Tollmien) (1962) 313--315.

\end{thebibliography}
\end{document}